\documentclass[aps,prl,twocolumn,footinbib,bibnotes,preprintnumbers,longbibliography]{revtex4-1}

\usepackage[utf8]{inputenc}
\usepackage{amsmath,amsfonts,epsfig,color,latexsym}
\usepackage{amssymb}
\usepackage[english, USenglish]{babel}
\usepackage{epsfig,psfrag}
\usepackage{slashed}
\usepackage[normalem]{ulem}

\usepackage{soul,xcolor}
\usepackage{hyperref}
\setstcolor{red}
\usepackage{comment}

\usepackage{subcaption}
\usepackage{multirow}

\usepackage{stackengine}
\usepackage{tikz}
\usetikzlibrary{positioning,arrows}
\usepgflibrary{arrows.meta}
\usetikzlibrary{decorations.pathmorphing}
\usetikzlibrary{decorations.markings}
\usetikzlibrary{math}
\usetikzlibrary{calc}

\tikzset{
Wilson_1/.style={double distance=1.1pt,postaction={decorate}, decoration={markings,mark=at position .08 with {\arrow{Stealth[scale=0.5]}},mark=at position .985 with {\arrow{Stealth[scale=0.5]}}}},
Scalar/.style={densely dashed},
Gauge/.style={decorate, draw=black, decoration={coil,aspect=0.5, post length = 0pt, pre length = 0pt, segment length=1.75pt,amplitude=1.4pt}},
Gauge-Back/.style={ultra thick, opacity=0.8, decorate, draw=white, decoration={coil,aspect=0.5, post length = 0pt, pre length = 0pt, segment length=1.75pt,amplitude=1.4pt}},
Fermion/.style={},
Fermion-Back/.style={ultra thick, opacity=0.8, draw=white}, 
}

\usepackage{enumitem}

\begin{document}
\newpage
\pagenumbering{arabic}

\newcommand{\yorgos}[1]{{\bf \textcolor{orange}{[#1 - GP]}}}
\newcommand{\be}{\begin{equation}}
\newcommand{\ee}{\end{equation}}
\newcommand{\norm}[1]{{\protect\normalsize{#1}}}
\newcommand{\p}[1]{(\ref{#1})}
\newcommand{\half}{\tfrac{1}{2}}
\newcommand \vev [1] {\langle{#1}\rangle}
\newcommand \ket [1] {|{#1}\rangle}
\newcommand \bra [1] {\langle {#1}|}
\newcommand \pd [1] {\frac{\pa}{\pa {#1}}}
\newcommand \ppd [2] {\frac{\pa^2}{\pa {#1} \pa{#2}}}
\newcommand{\ed}[1]{{\color{red} {#1}}}

\newcommand{\cI}{{\cal I}}
\newcommand{\cM}{{\cal M}} 
\newcommand{\cR}{{\cal R}} 
\newcommand{\cS}{{\cal S}} 
\newcommand{\cK}{{\cal K}}
\newcommand{\cL}{{\cal L}} 
\newcommand{\cF}{{\cal F}}
\newcommand{\cN}{{\cal N}}
\newcommand{\cA}{{\cal A}}
\newcommand{\cB}{{\cal B}}
\newcommand{\cG}{{\cal G}}
\newcommand{\cO}{{\cal O}}
\newcommand{\cY}{{\cal Y}}
\newcommand{\cX}{{\cal X}}
\newcommand{\cT}{{\cal T}}
\newcommand{\cW}{{\cal W}}
\newcommand{\cP}{{\cal P}}
\newcommand{\bP}{{\bar\Phi}}
\newcommand{\mK}{{\mathbb K}}
\newcommand{\nt}{\notag\\} 
\newcommand{\pa}{\partial}
\newcommand{\ep}{\epsilon}
\newcommand{\om}{\omega}
\newcommand{\bom}{\bar\omega}
\newcommand{\etap}{\bar\epsilon}
\newcommand{\vep}{\varepsilon}
\renewcommand{\a}{\alpha}
\renewcommand{\b}{\beta}
\newcommand{\g}{\gamma}
\newcommand{\s}{\sigma}
\newcommand{\la}{\lambda}
\newcommand{\tl}{\tilde\lambda}
\newcommand{\tm}{\tilde\mu}
\newcommand{\tk}{\tilde k}
\newcommand{\da}{{\dot\alpha}}
\newcommand{\db}{{\dot\beta}}
\newcommand{\dg}{{\dot\gamma}}
\newcommand{\dd}{{\dot\delta}}
\newcommand{\q}{\theta}
\newcommand{\bq}{{\bar\theta}}
\renewcommand{\r}{\rho}
\newcommand{\br}{\bar\rho}
\newcommand{\bQ}{\bar Q}
\newcommand{\bx}{\bar \xi}
\newcommand{\tx}{\tilde{x}}
\newcommand{\tr}{\mbox{tr}}
\newcommand{\+}{{\dt+}}
\renewcommand{\-}{{\dt-}}
\newcommand{\ti}{{\textup{i}}}

\newcommand{\dlog}{d{\rm log}}
\newcommand{\tred}[1]{\textcolor{red}{\bfseries #1}}
\newcommand{\eps}{\epsilon}

\newcommand{\api}{{\left(\frac{\alpha}{\pi}\right)}}

\newcommand{\hexagon}{\rm hex}

\preprint{DESY 20-204}
\preprint{MPP-2020-216}
\title{
Cluster algebras for Feynman integrals
}

\author{Dmitry\ Chicherin}
\email{chicheri@mpp.mpg.de}
\affiliation{Max-Planck-Institut f{\"u}r Physik, Werner-Heisenberg-Institut, 80805 M{\"u}nchen, Germany}

\author{Johannes\ M.\ Henn}
\email{henn@mpp.mpg.de}
\affiliation{Max-Planck-Institut f{\"u}r Physik, Werner-Heisenberg-Institut, 80805 M{\"u}nchen, Germany}

\author{Georgios\ Papathanasiou}
\email{georgios.papathanasiou@desy.de}
\affiliation{DESY Theory Group, DESY Hamburg, Notkestrasse 85, 22607 Hamburg, Germany}

\begin{abstract}
We initiate the study of cluster algebras in Feynman integrals in dimensional regularization. We provide evidence that four-point Feynman integrals with one off-shell leg are described by a $C_{2}$ cluster algebra, and we find cluster adjacency relations that restrict the allowed function space. By embedding $C_{2}$ inside the $A_3$ cluster algebra, we identify these adjacencies with the extended Steinmann relations for six-particle massless scattering. The cluster algebra connection we find restricts the functions space for vector boson or Higgs plus jet amplitudes, and for form factors recently considered in $\mathcal{N}=4$ super Yang-Mills. 
We explain general procedures for studying relationships between alphabets of generalized polylogarithmic functions and cluster algebras, and use them to provide various identifications of one-loop alphabets with cluster algebras.
In particular, we show how one can obtain one-loop alphabets for five-particle scattering from a recently discussed dual conformal eight-particle alphabet related to the $G(4,8)$ cluster algebra.
\end{abstract}

\maketitle

\section{Introduction}

Recent years have seen the emergence of unexpected mathematical structures in scattering amplitudes,
especially so in $\mathcal{N}=4$ supersymmetric Yang Mills (sYM). 
Planar loop integrands in this theory are in principle known to all loop orders \cite{ArkaniHamed:2010kv}, with the on-shell data entering the former being described by a positive Grassmannian \cite{ArkaniHamed:2009dn}. Furthermore the integrand has a dual geometric description \cite{Arkani-Hamed:2013jha}.
There is considerable evidence that cluster algebras play an important role for the amplitudes, both at the level of the loop integrand \cite{Arkani-Hamed:2016byb}, as well as for the functions obtained after integration \cite{Golden:2013xva}. 
For example, planar six- and seven-gluon scattering amplitudes appear to be governed by the finite $A_3$ and $E_6$ cluster algebras, respectively. 
This suggests that their function space is a certain set of generalized polylogarithms, which is the starting point for the bootstrap program \cite{Dixon:2011pw,Drummond:2014ffa}. 
Further constraints  
come from the absence of discontinuities in overlapping channels, the (extended) Steinmann relations \cite{Caron-Huot:2016owq,Caron-Huot:2018dsv,Caron-Huot:2019bsq}, which are closely related to cluster adjacency properties \cite{Drummond:2017ssj,Drummond:2018dfd}.
These findings have been instrumental for bootstrapping amplitudes to very high loop orders, see e.g. \cite{Dixon:2016nkn,Drummond:2018caf,Caron-Huot:2019vjl,Dixon:2020cnr}, and the review \cite{Caron-Huot:2020bkp}.

How general is the appearance of cluster algebras in quantum field theory?
On the one hand, all known cases are related to planarity, and concern finite parts of amplitudes in $\mathcal{N}=4$ sYM, which have additional symmetries \cite{Drummond:2008vq,Berkovits:2008ic,Drummond:2009fd}.
On the other hand, it motivates us that 
 several 
 structures initially found in $\mathcal{N}=4$ sYM, such as insights into the transcendental structure of Feynman integrals \cite{ArkaniHamed:2010gh}, their evaluation \cite{Henn:2013pwa} and dealing with their analytic properties and identities between them \cite{Goncharov:2010jf}, are by now common tools in generic quantum field theories, as reviewed in \cite{Duhr:2019wtr,Henn:2020omi}.
In this Letter we initiate a study of cluster algebras in Feynman integrals in $D=4- 2 \eps$ dimensions, without relying on extra symmetries.

\section{Cluster algebras and associated function spaces}

Cluster algebras \cite{1021.16017,1054.17024,CAIII,CAIV} (see  \cite{2008arXiv0807.1960K,Lampe13,Fomin2016,Fomin2017} for introductory articles) are commutative algebras equipped with a distinguished set of generators $a_i$, the \emph{cluster $\mathcal{A}$-coordinates}, grouped into overlapping subsets $\boldsymbol{a}\equiv\{a_1,\ldots,a_d\}$ of \emph{rank} $d$,  the \emph{clusters}. Starting from an initial cluster, they may be constructed recursively by a \emph{mutation} operation on the cluster coordinates. They may also be generalized to contain \emph{frozen coordinates} or \emph{coefficients} $\{a_{d+1},\ldots,a_{d+m}\}$, whose main difference from the cluster coordinates is that they do not mutate.

How the cluster coordinates transform under a mutation is encoded in an integer $(d+m)\times d$ \emph{exchange matrix} $B$, whose components we will denote as $b_{ij}$. Restricting to the components with $i,j\le d$ corresponds to the \emph{principal part} of $B$, which must be skew-symmetrizable. 
Then, mutating a cluster $(\boldsymbol{a},B)$ along the $k$-th variable, 
with $1 \le k\le d$, we obtain the new cluster $(\boldsymbol{a}',B')$, whose exchange matrix $B'$ is related to the previous one by
\begin{align}\label{eq:Bmutation}
	b'_{ij} = 
		\begin{cases}
			-b_{ij} \; &\text{for}\,i=k\,\text{or}\,j=k\\
			b_{ij} + \left[-b_{ik}\right]_+b_{kj} + b_{ik}\left[b_{kj}\right]_+\; &\text{otherwise}
		\end{cases}\,,
\end{align}
where $\left[x\right]_+ = \max\left(0,x\right)$. The cluster coordinates $a_{i}$ are unchanged for $i\neq k$ and $a_{k}$ is mutated according to
\begin{equation}
	\label{equ:clusterMutation}
	a'_k = a_k^{-1}\left(\prod_{i=1}^{d+m}a_{i}^{\left[b_{ik}\right]_+} + \prod_{i=1}^{d+m}a_{i}^{\left[-b_{ik}\right]_+}\right) \,.
\end{equation}

\begin{figure}
\centering
\begin{subfigure}{0.45\textwidth}
  \includegraphics[width=\textwidth]{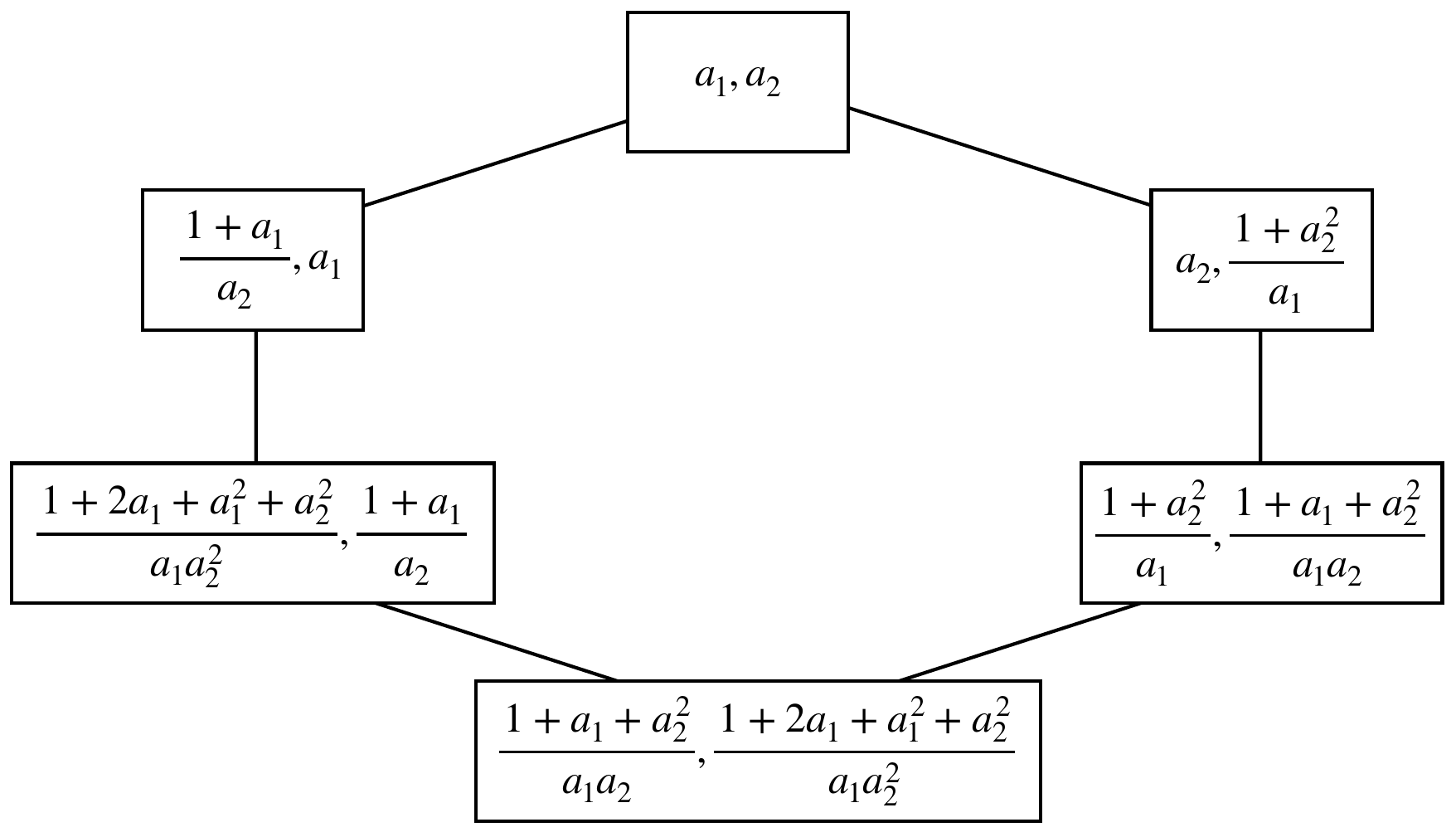}
\end{subfigure} 
 \caption{The exchange graph of the $C_2$ cluster algebra, with cluster coordinates ordered as $a_i,a_{i+1}$.
 }
 \label{B2C2ExchangeGraph}
 \end{figure}

The $C_2$ cluster algebra will play a prominent role in this Letter. For the coefficient-free case $m=0$ it is defined by
\be
B=\left(
\begin{array}{cc}
0&1\\
-2&0
\end{array}\right)\,.
\ee
Then, it is easy to show that under a mutation 
\eqref{eq:Bmutation} $B$ only changes by a sign, and thus the corresponding cluster transformation~\eqref{equ:clusterMutation} simplifies to
\be\label{simplifiedmutationrule}
a_{m+1}a_{m-1}=
\begin{cases}
1+a_m&\text{if $m$ is odd}\,,\\
1+a_m^2&\text{if $m$ is even}\,,\\
\end{cases} 
\ee
where $\{a_1,a_2\}$ are the $\mathcal{A}$-coordinates of the initial cluster, $a_3=a'_1$ and so on. The coordinates obtained in this manner are shown in the \emph{exchange graph} of Fig.~\ref{B2C2ExchangeGraph}, where clusters $\{a_i,a_{i+1}\}$ are represented by vertices, and the mutations relating them by edges. The circle topology indicates that mutating six times takes us back to where we started, $a_{i+6}=a_i$.

Below we will also find it useful to consider an alternative set of coordinates, 
called {\it cluster $\mathcal{X}$-coordinates} \cite{FG03b}, which are defined as
\begin{equation}
	\label{eq:xtoa}
	x_{i} \equiv \prod_{l=1}^{d+m}a_{l}^{b_{li}}\,,\quad i=1,\ldots d\,.
\end{equation}
The $\mathcal{A}$-coordinate mutation \eqref{equ:clusterMutation} and eq.~\eqref{eq:xtoa} imply a relation between the associated $\mathcal{X}$-coordinates, which may be written directly as
\begin{equation}\label{eq:xMutation}
x_i' = \begin{cases}
    1/x_i              &   k = i\,,\\
    x_i\bigl(1+x_k^{-{\rm sgn}(b_{ki})}\bigr)^{-b_{ki}} & k \neq i\,.
  \end{cases}
\end{equation}
In fact the latter provide another way of defining cluster algebras (more precisely, cluster Poisson varieties): one may start with the $\mathcal{X}$-coordinates and the principal part of $B$ of the initial cluster, and obtain all other clusters by virtue of the $\mathcal{X}$-coordinate mutation \eqref{eq:xMutation}.

Let us now associate a natural function space to a cluster algebra, given a set $\{a_i\}$ of $\mathcal{A}$- (or similarly, $\mathcal{X}$-) coordinates.
A {\it cluster (polylogarithm) function} $f$ \cite{Parker:2015cia} of (transcendental) weight $w$ has the defining property that its differential has the form
\begin{align}\label{definition_cluster_function}
d \,f^{(w)} = \sum_{i} f^{(w-1)}_{i} d \log a_{i} \,,
\end{align}
where the $f_{i}$ are again cluster functions, of weight ${(w-1)}$. The iterative definition starts with the weight zero function, which is a constant.
From this it follows that cluster functions of weight $w$ can be expressed as $w$-fold Chen iterated integrals \cite{chen1977}.

The definition (\ref{definition_cluster_function}) is very similar to {\it{canonical differential equations}} satisfied by certain classes of Feynman integrals \cite{Henn:2013pwa},
\begin{align}\label{canonicalDE}
d\, {\bf f}(\vec{z};\eps)  = \eps  \left[ \sum_i  {\bf A}_{i} d \log \alpha_i(\vec{z}) \right] {\bf f}(\vec{z};\eps) \,,
\end{align}
where ${\bf f}$ is a basis of Feynman integrals under consideration, ${\bf A}_{i}$ are constant matrices, and the $\alpha_i$ are algebraic functions of the kinematic variables $\vec{z}$, and $d= \sum_j dz_j  \partial_{z_j}$. Solving eq. (\ref{canonicalDE}) as a series in $\eps$ yields Chen iterated integrals, with the weight corresponding to the order in $\eps$.  
Moreover, the {\it symbol} \cite{Goncharov:2010jf} of the answer, which amounts to the solution modulo integration constants, can be read off from eq. (\ref{canonicalDE}).
The set of $\alpha$'s is called the  {\it{alphabet}}, and its elements are called {\it letters}.

Knowing the alphabet of a given Feynman integral or scattering amplitude is an important piece of information, because it is essential for bootstrapping the answer.
In the following, we wish to study whether the alphabets of certain Feynman integrals coincide with the $\mathcal{A}$- or $\mathcal{X}$-coordinates of some cluster algebra (up to multiplicative redefinitions).

\section{$C_{2}$ cluster algebra and four-particle scattering with one off-shell leg}
\begin{figure}
\centering
\begin{subfigure}{.15\textwidth}
   \includegraphics[width=\textwidth]{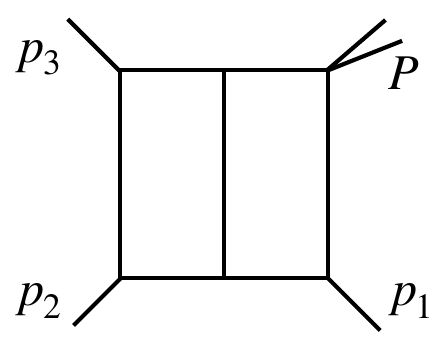}
\end{subfigure}%
\;\;
\,\,\begin{subfigure}{.15\textwidth}
 \includegraphics[width=\textwidth]{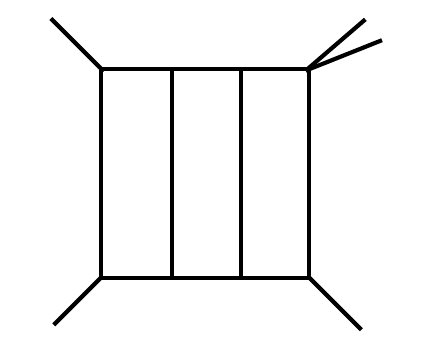}
\end{subfigure}%
\,\,\begin{subfigure}{.15\textwidth}
 \includegraphics[width=\textwidth]{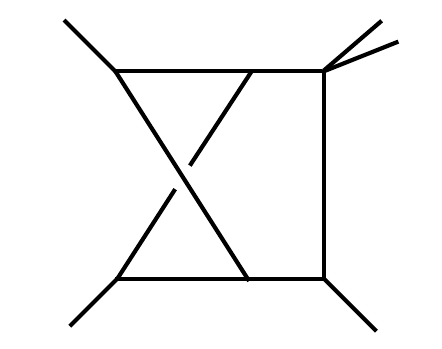}
\end{subfigure}%
\caption{Examples of known two- and three-loop four-point integrals with one off-shell leg, $P^2 \neq 0$.}
\label{fig:2Loop1MassBoxes}
\end{figure}

We consider four-particle scattering processes with one off-shell leg and massless internal lines, cf. Fig.~\ref{fig:2Loop1MassBoxes}. 
This kinematic situation has many important physical applications. 
It applies to helicity amplitudes of a Higgs and three partons \cite{Gehrmann:2011aa,Duhr:2012fh} (in the limit of infinite top quark mass),
or of a vector boson and three partons \cite{Gehrmann:2013vga}, that are relevant for collider physics. 
Interestingly, it has been observed that the Higgs amplitudes bear close resemblance to form factors of a composite operator inserted into three on-shell states in $\mathcal{N}=4$ sYM \cite{Brandhuber:2017bkg}.
What is more, very recently, an integrability description for form factors in $\mathcal{N}=4$ sYM was derived \cite{Sever:2020jjx}, which provides valuable all-loop and even non-perturbative information.

Denoting the Lorentz invariants by $z_1 =s / P^2$ and $ z_2=  t/ P^2$, with $s = 2 p_1 \cdot p_2$ and $t=2 p_2 \cdot p_3$, 
we note that all known Feynman integrals (i.e. all planar and non-planar two-loop integrals \cite{Gehrmann:2000zt,Gehrmann:2001ck}, and certain planar three-loop integrals \cite{DiVita:2014pza}, see Fig.~\ref{fig:2Loop1MassBoxes}) in this kinematics can be expressed in terms of the following alphabet (to all orders in $\eps$),
\begin{align}\label{eq:2dHPL}
\Phi_{\text{2dHPL}}=\{z_1,z_2, z_3, 1-z_1,1-z_2,1-z_3\}\,,
\end{align}
with $z_1 + z_2 + z_3 =1$.
The associated class of functions, dubbed two-dimensional harmonic polylogarithms (2dHPL), is well-studied in the physics literature \cite{Gehrmann:2001jv,Duhr:2012fh}
\footnote{At one loop, the five-letter subalphabet $\{z_1,z_2,1-z_1,1-z_2,z_1+z_2\}$ is sufficient. It is well-known that it corresponds to the $A_{2}$ cluster algebra.
Closely related function spaces have also appeared in off-shell form factors, four-point CFT correlation functions, and in the soft anomalous dimension matrix.}.

\renewcommand{\arraystretch}{1.25}
\begin{table}[tp]
\caption{Dimension of the $C_2$ cluster functions space (modulo transcendental constants), after constraints.}
\begin{tabular}{|c| c c c c c c c  c |}
\hline
weight  & 1  & 2 & 3 & 4 & 5 & 6 & 7 & 8 \\
\hline
{{First entry condition}}
 & 3  & 12 & 45 & 165 & 597 & 2143& 7653& 27241\\ 
Adjacency constraint & 3  & 12 & 42 & 138 & 438 & 1362 & 4182& 12738 \\ 
\hline
\end{tabular}
\label{tableB2integrability}
\end{table}%

We now show that surprisingly, the 2dHPL alphabet (\ref{eq:2dHPL}) is equivalent to the $C_{2}$ alphabet given in Fig.~\ref{B2C2ExchangeGraph}.
To see this one applies the mapping 
\begin{equation}
z_1 = - \frac{a_2^2}{1+a_1} \,, \quad z_2= - \frac{1+a_1+a_2^2}{a_1 (1+a_1)}  \,. \label{z1z2a1a2}
\end{equation}
This identification allows us to make a number of interesting observations:

1. The cluster $\mathcal{X}$-coordinates are useful for choosing a particular polylogarithmic representation of the functions and for writing down functional identities \cite{Gehrmann:2001jv,Duhr:2012fh,Parker:2015cia,Golden:2014xqa}. For example, if $x$ is an $\mathcal{X}$-coordinate, then ${\rm Li}_{n}(-x)$ is a cluster function, as can be understood from Eq.(\ref{eq:xMutation}). In \cite{Parker:2015cia} an algorithm for constructing a basis of $A_n$ cluster functions at any weight is given. 
Moreover, we note that the $C_{2}$ cluster mutations generate a subset of the automorphisms of the 2dHPL alphabet.

2. The cluster algebra suggests to group the alphabet letters according to the clusters. 
In certain functions in $\mathcal{N}=4$ sYM the interesting property of {\it{cluster adjacency}} \cite{Drummond:2017ssj} has been observed, which in the language of differential equations \eqref{canonicalDE} translates to the statement that ${\bf A}_{i}.{\bf A}_{j} = 0$ if $\alpha_i,\alpha_j$ do not appear together in any cluster. 
Note that this proves the adjacency property to all orders in $\eps$.
Inspecting the differential equations for the two- and three-loop planar integrals of \cite{Gehrmann:2000zt,DiVita:2014pza}, we observe that they obey a subset of $C_{2}$ adjacency conditions, namely
\be\label{eq:ObservedAdjacency}
{\bf A}_{i}.{\bf A}_{j} = 0\,, \;\; {\rm for} \;  i,j \in \{1,3,5\} \; {\rm with} \; i \neq j\,.
\ee
{Eqs. \eqref{eq:ObservedAdjacency} also hold for the non-planar integrals of \cite{Gehrmann:2001ck}, as was first noticed in \cite{Dixon:2020bbt}.}
In terms of the equivalent alphabet (\ref{eq:2dHPL}),  this implies that the letters $1-z_i$ and $1-z_j$ for $i\neq j$ never appear next to each other in a symbol.
The adjacency conditions we observe significantly reduce the dimension of the space containing the aforementioned integrals, as shown in table \ref{tableB2integrability}. 

Furthermore, the adjacency property extendeds to rational functions \cite{Drummond:2018dfd},
as can be seen by inspecting for example the two-loop amplitudes of \cite{Brandhuber:2017bkg,Jin:2018fak}.
This is obvious because the latter only contain poles at $z_{i}=0$.

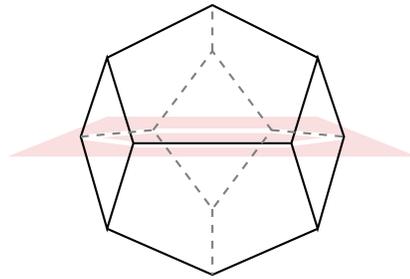
\begin{figure}
\begin{center}
\begin{tikzpicture}[scale=0.35]
\definecolor{bg}{RGB}{246,202,203}
\coordinate (A) at (1,0.8);
\coordinate (B) at (9,0.8);
\coordinate (C) at (13,-0.8);
\coordinate (D) at (-3,-0.8);
\draw[thick, color=white, fill=bg,fill opacity=0.6] (A) -- (B) -- (C) -- (D) -- cycle;

\draw[very thick,white] (0,0) -- (2.75,0.25) ;
\draw[very thick,white]  (7.25,0.25) -- (10,0);
\draw[very thick,white] (2,-0.25) -- (8,-0.25);
\draw[very thick,white] (0,0) -- (2,-0.25);
\draw[very thick,white] (8,-0.25) -- (10,0);
\draw[very thick,white] (2.75,0.25) -- (7.25,0.25);

\draw[very thick,white] (0,0-0.05) -- (2.75,0.25-0.05) ;
\draw[very thick,white]  (7.25,0.25-0.05) -- (10,0-0.05);
\draw[very thick,white] (2,-0.25-0.1) -- (8,-0.25-0.1);
\draw[very thick,white] (0,0-0.05) -- (2,-0.25-0.05);
\draw[very thick,white] (8,-0.25-0.05) -- (10,0-0.05);
\draw[very thick,white] (2.75,0.25-0.05) -- (7.25,0.25-0.05);

\draw[thick,fill=none] (0,0) -- (1,3) -- (2,-0.25) -- (1,-3.5)  -- cycle;
  \draw[thick,fill=none] (2,-0.25) -- (8,-0.25);
  \draw[thick,fill=none] (10,0) -- (9,3) -- (8,-0.25) -- (9,-3.5)  -- cycle;
  \draw[thick,fill=none] (1,3) -- (5,5) -- (9,3);
  \draw[thick,fill=none] (1,-3.5) -- (5,-5.25) -- (9,-3.5);
  \draw[thick,gray,dashed] (0,0.0) -- (2.75,0.25) ;
  \draw[thick,gray,dashed]  (7.25,0.25) -- (10,0.0);
  \draw[thick,gray,dashed]  (2.75,0.25) -- (5,3.25) -- (7.25,0.25) -- (5,-2.75) -- cycle;
  \draw[thick,gray,dashed]  (5,3.25) -- (5,5);
  \draw[thick,gray,dashed]  (5,-2.75) -- (5,-5.25);

 \end{tikzpicture}
 \end{center}
 \caption{
 The intersection of the exchange graph of the $A_3$ cluster algebra with the parity invariant plane (in pink) is shown in white. It can be identified with the $C_2$ cluster exchange graph shown in Fig.~\ref{B2C2ExchangeGraph}.} 
 \label{Stasheff_triang}
 \end{figure}

3. 
We can interpret the observed adjacency in the following way:  As shown in Fig.~\ref{Stasheff_triang}, $C_2$ is the parity-invariant surface of the $A_{3}$ cluster algebra, relevant for six-gluon amplitudes in $\mathcal{N}=4$ sYM 
\footnote{More generally, we find that $C_{n}$ can be identified as the parity-invariant surface inside $A_{2n-1}$, and that the same holds true for $F_4$ inside $E_{6}$. Similarly, we have obtained $G_{2}$ and $B_{n}$ from $D_{4}$ and $D_{2n-1}$, respectively. See also \cite{Arkani-Hamed:2020tuz}.}.
On the latter surface, corresponding to $\Delta\equiv(u+v+w-1)^2-4 u v w=0$, where $u,v,w$ are the dual conformal cross-ratios parametrizing the kinematics, only parity-even combinations of the $A_3$ $\cA$-coordinates are relevant. Choosing them as
\be
\begin{aligned}\label{eq:B2toA3Letters}
a_1&=\sqrt \frac{u}{v w} \,,\;\;\;\;\, a_3=\sqrt \frac{w}{u v} \,,\;\;\;\;\; a_5=\sqrt \frac{v}{u w}  \,,\\
 a_2&=\sqrt \frac{1-v}{v} \,,\;  a_4=\sqrt \frac{1-u}{u} \,,\; a_6=\sqrt  \frac{1-w}{w}  \,,
\end{aligned}
\ee
where $\{a_1,a_2\}$ are associated to the leftmost equatorial cluster in Fig.~\ref{Stasheff_triang},
$\{a_6,a_1\}$ to the left front cluster, and so on, we find that they satisfy exactly the $C_{2}$ mutation rule, eq. (\ref{simplifiedmutationrule}), on the $\Delta=0$ surface. 
By virtue of eq. (\ref{eq:B2toA3Letters}), we recognize that the adjacency restrictions~(\ref{eq:ObservedAdjacency}) precisely correspond to the extended Steinmann relations~\cite{Caron-Huot:2018dsv,Caron-Huot:2019bsq} (i.e. applying to any consecutive entries in the symbol) for six-particle 
massless scattering!\footnote{Note however that the first entries are $z_{i}$ with $i=1,2,3$, while the adjacency conditions apply to the $1-z_{i}$ entries, i.e. the relations found cannot be interpreted as Steinmann relations in the kinematic space of four-particle scattering with one off-shell leg.}\footnote{It is very interesting to note that the $A_{3}$ mutations that lie on the parity-invariant surface relate
$a_{2i-1}\leftrightarrow a_{2i+1}$, i.e. exactly the pairs appearing in the adjacency relations~\eqref{eq:ObservedAdjacency} we observe.}

4. Finally, let us comment on the massless limit $P^2 \to 0$. Denoting $z=t/s$, Eq. (\ref{eq:2dHPL}) reduces to the alphabet $\{z, 1+z \}$ in the limit.
This is consistent with the fact that all currently known Feynman integrals in this kinematics (up to three loops, planar and non-planar) satisfy differential equations (\ref{canonicalDE}) with this alphabet \cite{Henn:2013fah,Henn:2013nsa,Henn:2020lye}. 
It is interesting to note that one can obtain this on-shell four-particle alphabet from an $A_{1}$ cluster algebra with one frozen variable.

\section{Matching function spaces and cluster algebras, and applications}

\begin{table}
\caption{One-loop integrals related to cluster algebras.}
\begin{tabular}{|c|c|c||c|}
\hline
\multirow{2}{*}{ Integral family } & {\# } &   {\# } & \multirow{2}{*}{ cluster algebra }  \\ & var. & letters &  \\
 \hline
\multirow{2}{*}{ 1-loop Bhabha scattering \cite{Henn:2013woa} } & \multirow{2}{*}{2} & \multirow{2}{*}{8} & $\subset A_3$ \\ & & & $\lim C_3$ \\
\hline
 2-mass-easy box & 3 & 7 (9) & $ \subset A_3$ \\
 \hline
\multirow{2}{*}{2-mass-hard box} & \multirow{2}{*}{3} & \multirow{2}{*}{10} & $\subset C_3$ \\ & & & $\lim D_4$ \\
 \hline
 3-mass box & 4 & 16 & $ \subset C_4$ \\ 
 \hline
\end{tabular}
\label{tableidentifications}
\end{table}

There are many other cases where Feynman integrals evaluate to generalized polylogarithmic functions that are characterized by an alphabet.
Are any of them associated to cluster algebras? If so, how can we find this relation?

We use the following procedure to find {{embeddings of alphabets into cluster algebras}}.
Let $\vec{\alpha}$ be an alphabet depending on $d$ variables $z_1,\ldots,z_d$ in a rational way.
We would like to express all letters $\vec \alpha$ in terms of the letters $\vec\beta$ of a candidate cluster alphabet (with more or equal number of letters),  
\begin{align}\label{transformation}
d\log(\alpha_i) = \sum_j n_{ij} d\log(\beta_{j}) \,,
\end{align}
where $n_{ij}$ are integers.
We proceed in the following steps. Usually it is possible to simplify the letters $\vec{\alpha}$ by a birational change of variables, such that $\{ z_i, 1+z_i\}_{i=1}^d$ are among the new set of letters. Next, we order the letters according to their `complexity', namely the number of $z$-variables they depend on.
Parsing though the transformations with small integer coefficients $n_{ij}$, we identify those which are consistent with the one-variable letters.
Then we further restrict the set of allowed transformations demanding that two-variable letters 
also factorize into the cluster letters, and so on for letters of higher complexity.
This yields a host of novel matchings, which are indicated in table~\ref{tableidentifications} with the $\subset$ sign.

Let us illustrate this for the two-mass easy box integral family with massive legs $p_1$ and $p_3$.
The corresponding 10-letter alphabet is given by \footnote{In fact, only 8 independent combinations of the letters appear in the one-loop differential equations.}
\begin{align}
\Phi_{2me} = & \{ s,t,p_1^2,p_3^2, s-p_1^2 , s-p_3^2, t-p_1^2 , t-p_3^2, \notag\\
& \quad s t - p_1^2 p_3^2  , s+t-p_1^2 - p_3^2 \} \,.
\end{align} 
The change of variables
\begin{align}
z_1 = - \frac{p_1^2}{s} ,\;
z_2 = - \frac{t}{p_3^2} ,\;
z_3 = \frac{t-p_1^2}{s-p_3^2} \,,
\end{align}
brings it to a simpler form,
\begin{align}
\Phi_{2me} \approx & \{ z_1 , z_2, z_3, 1+z_1 , 1+z_2, 1+z_3 , \notag\\
&  \quad z_1 - z_2 , z_1 - z_3, z_2 - z_3 \}\,,
\end{align}
where we dropped a trivial mass scale. This 9-letter alphabet is equivalent to the cluster $A_3$ alphabet.

We can find further identifications by considering {\it degenerations} of cluster algebra alphabets with fewer variables.
What we mean by this is to consider the subalphabet obtained at a hypersurface $\alpha_{i}=0$ (for some $i$).
We have already seen an example of this, when going from $C_{2}$ to $A_{1}$. 
We note that another degeneration of $C_{2}$ yields the important alphabet $\{z, 1+z,1-z \}$ corresponding to harmonic polylogarithms \cite{Remiddi:1999ew}.
The results of our heuristic search with this method are denoted in table~\ref{tableidentifications} by `lim'.

\begin{table}
\caption{Six-dimensional integrals and cluster algebras.}
\begin{tabular}{|c|c|c||c|}
\hline
\multirow{2}{*}{ Integral  } & {\# } &   {\# } & \multirow{2}{*}{ cluster algebra }  \\ & var. & letters &  \\
 \hline
 1-mass hexagon & 4 & 16 &  $D_4$\\
 \hline
 2-mass-easy hexagon & 5 & 24 &  $\subset D_5$\\
 \hline
{  2-mass-hard hexagon } & \multirow{2}{*}{5} & \multirow{2}{*}{27} & \multirow{2}{*}{$\lim G(4,8)$} \\ pentagon with one off-shell leg & & &  \\
\hline
\end{tabular}
\label{tableidentifications2}
\end{table}

The identifications of alphabets discussed so far are valid to all orders in the dimensional regularization parameter $\eps$. 
However, when computing finite physical quantities, often simplifications occur, and a reduced alphabet is sufficient to describe the answer.
Moreover, it could be that cluster algebras appear in general only for integer dimensions.
For these reasons we find it is interesting to look for cluster algebras for finite Feynman integrals, or when truncating the $\eps$ expansion. 
It is known that the (dual conformal) six-dimensional hexagon integral corresponds to the $A_{3}$ cluster algebra. Here find that its one-mass version \cite{DelDuca:2011jm,DelDuca:2011wh}, which has $20$ letters and $4$ dimensionless variables, corresponds to the $D_{4}$ cluster algebra.
Likewise, we find that a `two-mass-easy' hexagon \cite{DelDuca:2011wh} can be embedded into the $D_{5}$ cluster algebra.

\begin{figure}[t]
\begin{align*}
\begin{array}{c}
\includegraphics[width=3.5cm]{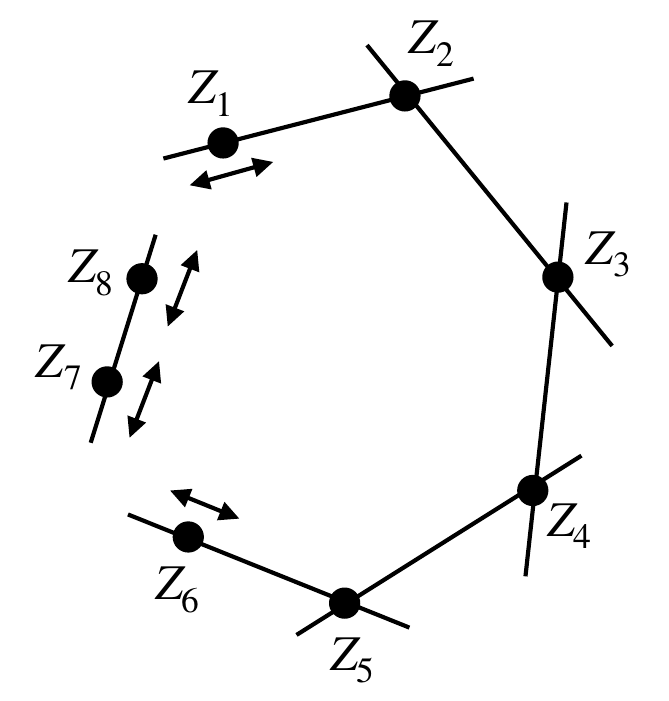}
\end{array} \qquad
\begin{array}{c}
\includegraphics[width=3.5cm]{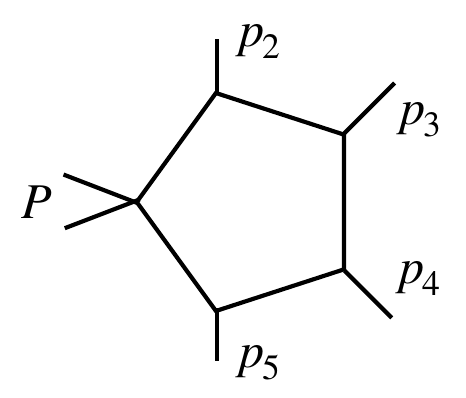}
\end{array}
\end{align*}
\caption{Momentum twistors in a dual conformal two-mass-hard hexagon configuration. 
This is equivalent to five-particle kinematics with one off-shell leg $P$.}
\label{two-mass-hard-hexagon}
\end{figure}

So far we have discussed finite cluster algebras, however in $\mathcal{N}=4$ sYM also infinite cluster algebras  are seen to play a role. 
For example, the infinite Grassmannian $G(4,8)$ algebra is expected to govern dual conformal eight-particle scattering, generalizing the six- and seven-particle case, with $A_{3} \sim G(4,6)$ and $E_{6} \sim G(4,7)$, respectively. Very recently, a natural way of identifying a finite set of letters from $G(4,8)$ has been proposed in \cite{Arkani-Hamed:2019rds,Henke:2019hve,Drummond:2019cxm}. 
Starting from the eight-particle alphabet thus obtained in the latter paper, here we will establish a connection with five-particle scattering in generic gauge theory. The eight-particle alphabet of \cite{Drummond:2019cxm} consists of 272 rational and 18 square-root letters, and contains both the alphabets of the  two-loop MHV~\cite{CaronHuot:2011ky} and NMHV amplitudes~\cite{Zhang:2019vnm} (see also \cite{He:2020vob,Mago:2020kmp,He:2020uhb}).
In a first step, we specify the generic eight momentum twistors \cite{Hodges:2009hk} such that they describe two adjacent massive legs, cf. Fig.~\ref{two-mass-hard-hexagon}. 
We find that this reduces the $G(4,8)$ alphabet to 30+5 rational and algebraic letters, consistent with the symbol of the six-dimensional two-mass-hard hexagon, as obtained from \cite{Spradlin:2011wp}.
The alphabet we find also applies to non dual-conformal integrals: by interpreting the bi-twistor $Z_{7} \wedge Z_{8}$ as the infinity twistor, the kinematics is equivalent to five-particle scattering with one massive leg, cf. Fig.~\ref{two-mass-hard-hexagon}.
Indeed we reproduce all letters appearing in the one-loop integrals of \cite{Abreu:2020jxa}, except for $W_{48}, W_{49}$. 
Taking in addition the {$P^2\to 0$} limit, we find $22$ letters from the planar pentagon alphabet \cite{Gehrmann:2015bfy}. 
These include all one-loop letters except for $W_{31}$ (using the notation of  \cite{Chicherin:2017dob}).
Moreover, the reduction of the octagon alphabet also captures some of the additional two-loop letters; upon cyclic symmetrization, we obtain all planar two-loop letters, apart from $W_{31}$.
This is very interesting, because $W_{31}$ has been observed to drop out of appropriately defined finite quantities, such as the hard part (after infrared subtraction) of the two-loop $\mathcal{N}=4$ sYM \cite{Abreu:2018aqd,Chicherin:2018yne} and $\mathcal{N}=8$ supergravity \cite{Chicherin:2019xeg,Abreu:2019rpt} amplitudes. 
The same is true for the hard part of the two-loop $q \bar{q} \to  \gamma\gamma\gamma$ amplitude \cite{Abreu:2020cwb,Vasilycomm} and the two-loop five-gluon amplitudes \cite{Badger:2018enw,Abreu:2019odu,Simonecomm} in QCD.

Interested experts may find the technical details of the identifications mentioned above in the Supplemental Material to this Letter.

\section{Discussion and outlook}

In this Letter, we have provided for the first time evidence that cluster algebras are relevant to Feynman integrals beyond $\mathcal{N}=4$ sYM. 

In particular, we have uncovered that the important class of two-dimensional harmonic polylogarithms corresponds to the $C_{2}$ cluster algebra. 
This suggests that a number of physical processes, including Higgs plus jet amplitudes in the heavy top quark limit, may be amenable to bootstrap methods \cite{Brandhuber:2012vm}.
Moreover, we found unexpected adjacency relations that further constrain the function space. By identifying $C_2$ as a subspace of the $A_3$ cluster algebra, we have shown that these adjacencies are equivalent to the extended Steinmann relations of planar six-particle scattering in $\mathcal{N}=4$ sYM~\cite{Caron-Huot:2018dsv,Caron-Huot:2019bsq}. 
Such relations have been used for six- and higher-particle amplitudes (and five-particle scattering with one off-shell leg \cite{Abreu:2020jxa}), 
but their implications are not yet known for massless five-particle scattering. This makes the adjacency relations  we found in a five-particle kinematics all the more interesting.
We have demonstrated that the constraints we found significantly reduce the relevant functions space, cf. Table~\ref{tableB2integrability}. This should prove valuable in bootstrap approaches. 
This is very timely in view of the recent integrability results for form factors in $\mathcal{N}=4$ sYM \cite{Sever:2020jjx}.

We have outlined heuristic procedures for matching Feynman integrals with cluster algebras, and found a host of one-loop identifications. This suggests a number of questions.
What other processes are related to cluster algebras? What cluster algebras describe the higher-loop versions of the cases discussed here? 
We leave these questions to future work, and hope that our results motivate a systematic taxonomy of the relationship between Feynman integrals and cluster algebras. 
 
In planar $\mathcal{N}=4$ sYM, there is a direct link between the kinematics and geometry of the scattering process, and cluster algebras.
This leads us to the most pressing open question: is there a similar story for generic Feynman integrals?

\acknowledgments

We thank S.~Badger, L. Dixon, P.~Mastrolia, A. McLeod,  V.~Sotnikov, M. Wilhelm and S.~Zoia 
for useful correspondence, N.~Henke for discussions, and M.~Spradlin for comments on the manuscript. JMH and GP thank the Higgs Centre for Theoretical Physics for hospitality during the workshop {\it Cluster Algebras and the Geometry of Scattering Amplitudes} in the early stages of this project. This research received funding from the European Research Council (ERC) under the European Union's Horizon 2020 research and innovation programme (grant agreement No 725110), {\it Novel structures in scattering amplitudes}. GP acknowledges support from the Deutsche Forschungsgemeinschaft under Germany’s Excellence Strategy EXC 2121 {\it Quantum Universe} 390833306.

\section{Supplemental material}

\subsection{Cluster algebras and one-loop alphabets}

Here we present the technical details of the identifications of one-loop alphabets with cluster algebras mentioned in the main part of the Letter.

In the following we deal with the cluster alphabets $\Phi$ of the $A_n, \, C_n, \, D_n$ series in their ${\cal X}$-coordinate formulation \p{eq:xMutation}, which contain ${n(n+3)}/{2}$, $n(n+1)$, and $n^2$ letters, respectively. We find convenient to make a birational change of variables to simplify the cluster letters \cite{Arkani-Hamed:2020tuz} as follows,

\begin{widetext}
\begin{align}
& \Phi_{A_n}= \bigcup\limits_{i=1}^n \{ z_i ,  1+ z_{i} \} \,  \cup  \, \bigcup\limits_{1\leq i<j \leq n} \{ z_i - z_j \} \; ,\qquad
\Phi_{C_n} = \Phi_{A_n} \cup \bigcup_{1\leq i \le j \leq n -1 } \{z_i z_j + z_n\} \label{Cn} \\
& \Phi_{D_n} = \bigcup\limits_{\substack{1\leq i< \\ j \leq n-2}} \{ z_i - z_j , z_i (z_{n-1}+z_n) - z_{n-1} z_n - z_i z_j  + z_i - z_j \} \,  \cup   \bigcup\limits_{i=1}^{n} \{ z_i , 1+ z_{i} \} \, \cup \bigcup\limits_{i=1}^{n-2} \{ z_i - z_{n-1}, z_i - z_{n}, z_i + z_{n-1} z_n \} \label{Dn}
\end{align}
\end{widetext}

{\it Two-mass hard box.} We consider the family of box integrals with two adjacent massive legs $p_1, \,p_2$. Parametrizing $p_1^2 = s y_1 y_2$ and $p_2^2 = s (1-y_1)(1-y_2)$, we find the following rational 11-letter alphabet,
\begin{align}
& \Phi_{2mh} = \{ s ,\, t ,\, y_1 ,\, 1 - y_1 ,\, y_2 ,\, 1 - y_2 ,\, y_1 - y_2 ,\, s y_1 y_2 - t , \notag \\
& s y_1(1-y_2) + t ,\, s y_2(1-y_1) + t ,\, s(1-y_1)(1-y_2) - t \} \,, \notag 
\end{align}
where $s = (p_1 + p_2)^2$ and $t = (p_2+ p_3)^2$.
After the birational change of variables 
\begin{align}
&t = -\frac{s z_1 z_3}{(1+z_2)(z_1 - z_3)} \notag\\
&y_1 = \frac{z_1}{z_1 - z_3} , \quad y_2 = \frac{1}{1+z_2} 
\end{align}
the alphabet becomes
\begin{align}
\Phi_{2mh} \approx \{ & s,z_1,z_2,z_3,1+z_1,1+z_2 ,1+z_3,\notag\\
&  z_1 - z_2, z_1 - z_3, z_2 - z_3, z_1 z_2 + z_3\} \,. \label{2mh_z}
\end{align}
Ignoring the trivial mass scale we establish the embedding in the $C_3$ cluster alphabet formulated as in eq. \p{Cn}, $\Phi_{2mh} \subset \Phi_{C_3}$.

Alternatively, one could start with the rank-4 cluster alphabet $D_4$, eq. \p{Dn}, and perform the degeneration $z_1 \to -1$,
\begin{align}
\lim_{z_1 \to -1} D_4 \approx \{ & z_2, z_3,z_4, 1+z_2 , 1+ z_3 , 1+z_4, \notag\\ 
& z_2 - z_3, z_2 - z_4, 1 - z_3 z_4 , z_2 + z_3 z_4  \}\,.
\end{align}
The resulting 10-letter alphabet is equivalent to \p{2mh_z} after a simple variable change,
\begin{align}
z_2 \to \frac{1}{z_1} ,\, z_3 \to \frac{1}{z_2} ,\, z_4 \to - \frac{z_3}{z_2} \,.
\end{align}

{\it Three-mass box.}
The family of box integrals with three massive legs $p_1,p_2,p_3$ is described by the 18-letter alphabet 
\begin{widetext}
\begin{align}
& \Phi_{3m} = \{ y_1 , y_2,y_3, y_4, 1- y_1 ,  1-y_2 ,  1- y_3 , 1-y_4 , y_1 - y_2 , y_3 - y_4 , 1-y_1 -y_3, 1-y_1 - y_4 ,  1-y_2 - y_3, 1-y_2 - y_4, \notag\\
& (1- y_1)(1- y_2) - y_3 y_4 , y_1 y_2 - (1-y_3)(1-y_4) , (1 - y_1 -y_2) (1-y_3)(1-y_4) + y_1 y_2 (1-y_3 - y_4), \, p_1^2 \}\,
\end{align}
\end{widetext}
in the rational parametrization of the kinematics
\begin{align}
& s = (1-y_1)(1-y_2) p_1^2 ,  \;\;
p_2^2 = y_1 y_2 p_1^2 , \notag\\
& t = (1-y_3)(1-y_4) p_1^2 ,  \;\;
p_3^2 = y_3 y_4 p_1^2 \,.
\end{align}
In fact, only 16 multiplicative combinations of the specified 17 dimensionless letters appear in the one-loop system of differential equations \p{canonicalDE}. After the change of variables
\begin{align}
& y_1 = \frac{z_1-z_2}{1+z_1},\quad
 y_2 = -\frac{z_4 + z_1 z_2}{z_1 - z_4},\notag\\
& y_3 = \frac{1+z_2}{1+z_3},\quad
 y_4 = \frac{z_3 (1+z_2)}{z_3 - z_4} 
\end{align}
and putting aside the trivial mass scale we find that $ \Phi_{3m} \subset \Phi_{C_4}$, cf. eq. \p{Cn}.

{\it One-loop Bhabha scattering.}
Bhabha scattering is a $2 \to 2$ process of massive electron scattering. The kinematics is specified by $s,t$ Mandelstam invariants and mass $m$. It is convenient to introduce dimensionless variables $x,y$: $-{s}/{m^2} = {(1-x)^2}/{x}$, $-{t}/{m^2} = {(1-y)^2}/{y}$. Then the relevant family of one-loop master integrals is described by the 8-letter alphabet \cite{Henn:2013woa} 
\begin{align}
\Phi_{\rm Bhabha} = \{ x, 1-x,1+x, y , 1-y,1+y, x+y,1+xy\} \,. \label{Bhabha} 
\end{align}
We can reproduce it by degenerating $z_3 \to -1$ the cluster alphabet $C_3$, eq. \p{Cn}, and identifying $x = z_1, \, y=z_2$. 

The same alphabet can be obtained from the $A_3$ cluster alphabet via non-singular degeneration. 
In order to see this it is more convenient to use the cluster ${\cal X}$-coordinates \p{eq:xtoa},
\begin{align}
\Phi_{A_3} \approx \{ & x_1, x_2, x_3, 1+ x_1 , 1+ x_2, 1+x_3 ,\notag\\
& 1+ x_2 + x_1 x_2, 1+ x_3 + x_2 x_3, 
,\notag\\
& 1+ x_3 + x_2 x_3 + x_1 x_2 x_3\}\,,
\end{align}
where $(x_1,x_2,x_3)$ is the initial seed with the exchange matrix $B$,
\begin{align}
B = \begin{pmatrix}
0 & -1 & 0 \\
1 & 0 & -1 \\
0 & 1 & 0
\end{pmatrix} .
\end{align}
Then the non-singular limit $x_1 \to - {1}/{x_3}$ and the following change of variables,
\begin{align}
x_3 = x , \quad x_2 = - \frac{x + y}{1-x} 
\end{align} 
bring the cluster alphabet $\Phi_{A_3}$ to the form of the one-loop Bhabha scattering alphabet \p{Bhabha}. Let us recall that the $A_3$ cluster alphabet describes the planar hexagon scattering in $\cN = 4$ sYM theory. The imposed constraint $x_1 x_3 = -1$ on the ${\cal X}$-variables corresponds to $u+v+w=1$ in the space of dual-conformal cross-ratios.

{\it Six-dimensional hexagons in easy-mass configurations.}
The six-dimensional hexagon integrals with up to three massive legs in the easy-mass configuration (where at least one massless leg separates two massive legs) are pure dual-conformal invariant functions of uniform weight three \cite{DelDuca:2011wh}. They depend on the cross-ratios $u_i$ in the dual momenta coordinates, with five cross-ratios in the two-mass-easy configuration and four-cross ratios in the one-mass configuration. The letters of the alphabets are algebraic in the cross-ratios, and in order to eliminate square roots it is very convenient to use a momentum-twistor parametrization of the cross-ratios. In order to make contact with notations in \cite{DelDuca:2011wh} we reproduce the rational parametrization of the cross-ratios $u_i$ by $y_1,\ldots,y_5$ in the two-mass-easy configuration (as compared to eq. (44) in \cite{DelDuca:2011wh}, we made one of the massive legs massless by putting there $y_8 \to 0$, i.e. $u_6 \to 0$, and we relabelled $x_5,x_8,y_2,z_2,z_5$ with $y_1,\ldots,y_5$, respectively), 
\begin{align}
&u_1 = \frac{y_4-y_5}{(1-y_3)(1-y_5)}\, , \,
u_2 = \frac{y_1 - y_2}{(1-y_2)(1-y_5)} \,, \notag\\
& u_3 = -\frac{y_3}{(1-y_2)(1-y_3)} \,,\,
u_4 = \frac{y_4 (1-y_5)}{y_4-y_5} \,, \notag\\
& u_5 = \frac{y_1 (1-y_2)}{y_1 - y_2} \,.
\end{align} 
We obtain the one-mass configuration from the two-mass-easy configuration by putting $y_1 \to 0$, which corresponds to $u_5 \to 0$.
Then the one-mass and two-mass-easy hexagon alphabets extracted from the one-loop calculation of \cite{DelDuca:2011wh} contain $16$ and $24$ letters, respectively, and they take the following form
\begin{widetext}
\begin{align}
\Phi_{\hexagon_{1m}} = \{ & y_2, y_3 , y_4 , y_5 , 1-y_2, 1-y_3 , 1-y_4 , 1-y_5, y_4 - y_5, 1-y_3 - y_4 , 1-y_2 + y_2 y_3 , 1-y_5 + y_2 y_5 , y_4 - y_5 + y_2 y_5 , \notag\\
&  1- y_3 - y_4 + y_3 y_5  , y_4 - y_5 + y_2 y_5 - y_2 y_3 y_5 , 1-y_2- y_4 + y_2 y_3 + y_2 y_4 \}\,, \notag \\[0.2cm]
\Phi_{\hexagon_{2me}} = \{ & y_1, y_2, y_3, y_4, y_5, 1-y_1, 1-y_2, 1-y_3, 1-y_4, 1-y_5, y_1 - y_2,  y_4 - y_5 , 1 -y_3 - y_4 , 1 - y_1 - y_5 , 1 -y_2 + y_2 y_3 , \notag\\ 
& y_1 - y_2 + y_2 y_3 , y_4 - y_5 - y_1 y_4 , 1- y_1 - y_5 + y_2 y_5 , 1- y_3 - y_4 + y_3 y_5 , y_4 - y_5 -y_1 y_4 + y_2 y_5, \notag\\
& 1-y_2 - y_4 + y_2 y_3 + y_2 y_4 , 1- y_1 - y_3 - y_4 + y_1 y_3 + y_1 y_4 + y_3 y_5 , y_4 - y_5 - y_1 y_4  + y_2 y_5 - y_2 y_3 y_5 , \notag\\
& y_1 - y_2 - y_1 y_4 + y_2 y_3 + y_2 y_4 - y_2 y_3 y_5 \}\,.
\end{align}
\end{widetext}
The birational change of variables 
\begin{align}
& y_1 = \frac{z_3 - z_5}{z_2 - z_5} \,,\,
y_2 = -\frac{z_3 - z_5}{(1+z_4)z_5} \,, \notag\\ 
&
y_3 = \frac{z_1 (1+z_5)}{z_1-z_5} \,,\,
y_4 = \frac{(z_1 - z_3)z_5}{z_3(z_1 - z_5)} \,,\notag\\
&
y_5 = \frac{(z_2 - z_3)z_5}{z_3(z_2 - z_5)} \,, \label{2mHexChange}
\end{align}
shows that $\Phi_{\hexagon_{2me}}$ is embedded in the 25-letter cluster alphabet $D_5$ in notations of eq. \p{Dn}, i.e. $\Phi_{\hexagon_{2me}} \subset \Phi_{D_5}$. More precisely, the extra letter of the cluster alphabet is $z_2 (z_4 + z_5) - z_4 z_5 - z_2 z_3 + z_2  - z_3$. 

The change of the variables \p{2mHexChange} relating the two-mass-easy hexagon alphabet and the cluster alphabet is compatible with the degenerating $y_1 \to 0$ into the one-mass hexagon that demands $z_2 \to \infty$. Indeed, the resulting birational transformation  
\begin{align}
& y_2 = -\frac{z_3 - z_5}{(1+z_4)z_5} \,, \,
y_3 = \frac{z_1 (1+z_5)}{z_1-z_5} \,, \notag\\  
&
y_4 = \frac{(z_1 - z_3)z_5}{z_3(z_1 - z_5)} \,, \, y_5 = \frac{(z_2 - z_3)z_5}{z_3(z_2 - z_5)} \,, \label{1mHexChange}
\end{align}
establishes the equivalence of the one-mass hexagon with the $D_4$ cluster alphabet $\Phi_{\hexagon_{1m}} \approx \Phi_{D_4}$. 


\subsection{Reduction of the octagon alphabet to pentagon functions}

\begin{figure}[t]
\includegraphics[width=3.5cm]{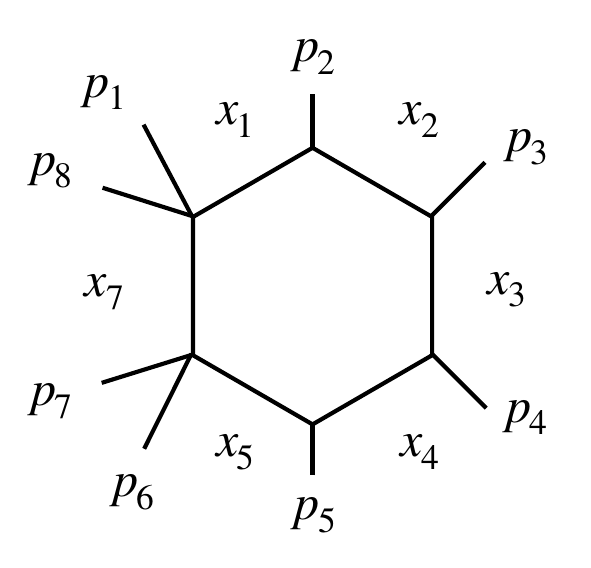}
\caption{Two-mass-hard hexagon configuration. The dual coordinates $x_{i}$ denote region momenta.}
\label{two-mass-hard-hexagon2}
\end{figure}

We consider an alphabet describing the dual-conformal invariant finite part part of the eight-particle amplitude in $\cN = 4$ sYM theory, and we show how to reduce it to the pentagon scattering with one massive leg. There are several alphabets relevant for eight-particle scattering in $\cN = 4$ sYM. In order to be specific, we consider the dual-conformal invariant alphabet \cite{Zhang:2019vnm} containing 172 rational letters and 18 algebraic letters in the momentum twistor variables. The latter involve two different square roots. We refer to it as the octagon alphabet $\Phi_{oct}$. The kinematics and the alphabet letters can be parametrized by 9 independent variables.

We introduce the dual momenta $x_i$ such that $x_{i+1} - x_{i} = p_{i+1}$  to describe the light-like kinematics $p_i^2 = 0$, $i=1,\ldots,8$, which automatically take into account momentum conservation. It is convenient to specify the kinematics by 8 unconstrained momentum twistors $Z_i$, which are points on $\mathbb{P}^3$, such that a pair of twistors (a line in $\mathbb{P}^3$) represents a dual momentum, $x_i \sim Z_i \wedge Z_{i+1}$. Intersecting lines $Z_i \wedge Z_{i+1}$ and $Z_{i+1}\wedge Z_{i+2}$ correspond to a pair of null-separated dual momenta, $(x_i-x_{i+1})^2 = 0$, and the intersection point $Z_{i+1}$ corresponds to the light-like momentum $p_{i+1}$. We would like that pairs $p_1$, $p_8$ and $p_6$, $p_7$ of the light-like momenta form massive corners (two-mass-hard dual-conformal hexagon), see Fig.~\ref{two-mass-hard-hexagon2}. It corresponds to the configuration of the momentum twistors depicted in Fig.~\ref{two-mass-hard-hexagon}. Twistor translations which do not alter the line configuration are allowed. We are interested in those multiplicative combinations of $\Phi_{oct}$ letters which are completely specified by this kinematic configuration, i.e. they are annihilated by four differential operators
\begin{align}
Z_5 \cdot \pa_{Z_6}  ,\, 
Z_2 \cdot \pa_{Z_1} ,\,
Z_7 \cdot \pa_{Z_8} , \,
Z_8 \cdot \pa_{Z_7} \,.
\end{align}
Resolving these constraints we obtain the 27-letter alphabet $\Phi_{\hexagon_{2m}}$ with 22 rational letters and 5 algebraic letters with one square root (in the momentum twistor variables). The letters of $\Phi_{\hexagon_{2m}}$ are functions of 5 independent variables. 

Switching to the frame $x_7 \to \infty$ in the dual-conformal letters of $\Phi_{\hexagon_{2m}}$ we obtain the one-mass pentagon configuration without any extra symmetry restrictions, cf. Fig.~\ref{two-mass-hard-hexagon}, where $P = p_6+p_7 + p_8 + p_1$, such that $P^2 \neq 0$. The latter is equivalent to identifying $Z_7 \wedge Z_8$ as an infinity bi-twistor that allows us to express `distance' $x_{ij}^2 = (x_i - x_j)^2$ in terms of the twistor 4-brackets $\vev{abcd} \equiv \det(Z_a Z_b Z_c Z_d)$, and eventually, transform the alphabet to the Mandelstam variables $s_{ij} = (p_i + p_j)^2$,
\begin{align}
& s_{12} = x_{25}^2 = \frac{\vev{2356}}{\vev{2378}\vev{5678}} \,,\, 
 s_{23} = x_{13}^2 = \frac{\vev{1234}}{\vev{1278}\vev{3478}} \,,\notag\\ 
& s_{34} = x_{24}^2 =  \frac{\vev{2345}}{\vev{2378}\vev{4578}} \,,\, 
 s_{45} = x_{35}^2 = \frac{\vev{3456}}{\vev{3478}\vev{5678}} \,,\notag\\
& s_{15} = x_{14}^2 = \frac{\vev{1245}}{\vev{1278}\vev{4578}} \,,\, 
 p_1^2 = x_{15}^2 = \frac{\vev{1256}}{\vev{1278}\vev{5678}} \,.
\end{align}
Thus, comparing the 28-letter alphabet (after restoring an overall scale) with the letters $\{ W_{i} \}_{i = 1}^{58}$ of the planar one-mass pentagon alphabet from \cite{Abreu:2020jxa}, we identify among the latter
\begin{align}
&
W_{1}, \ldots, W_{9}, W_{12}, \ldots, W_{15}, W_{18}, W_{19}, W_{22}, W_{23},W_{24},
\notag\\
& W_{33} , W_{34}, W_{37}, W_{38}, W_{40}, W_{43}, \ldots , W_{47}\,, \label{one-loop-pent-1m}
\end{align}
and the square-root inherited from the algebraic letters of $\Phi_{\hexagon_{2m}}$ equals to $\sqrt{\Delta_3}$ of \cite{Abreu:2020jxa}, which is the typical square root of the three-mass-triangle Feynman integral with massive corners $P, p_2+p_3,p_4+p_5$. This alphabet is enough to express the finite one-loop one-mass-pentagon Feynman integrals as well as finite parts of the one-loop scattering amplitudes (hard functions).

Larger octagon alphabets are also discussed in the literature. In particular, the parity restricted tropical ${\rm G}(4,8)$ alphabet presented in \cite{Drummond:2019cxm} contains 272 rational and the same 18 algebraic letters as $\Phi_{oct}$. The previous reduction procedure in this case leads to a bigger set of the planar one-mass pentagon letters \cite{Abreu:2020jxa}. In addition to the one-loop letters \p{one-loop-pent-1m}, it also captures genuine two-loop letters
\begin{align}
W_{16},W_{17},W_{29},W_{30},W_{52},W_{53},W_{55},W_{56\,}.
\end{align}    
Let us note that after taking the soft limit $P^2 \to 0$ and doing cyclic symmetrization (or equivalently, doing cyclically shifted reductions of the octagon alphabet) we reproduce all letters of the planar two-loop massless pentagon alphabet \cite{Chicherin:2017dob} which are needed to express the finite part of the scattering amplitudes (hard functions).


 \bibliographystyle{apsrev4-1} 

\bibliography{BibFile}

\begin{thebibliography}{84}%
\makeatletter
\providecommand \@ifxundefined [1]{%
 \@ifx{#1\undefined}
}%
\providecommand \@ifnum [1]{%
 \ifnum #1\expandafter \@firstoftwo
 \else \expandafter \@secondoftwo
 \fi
}%
\providecommand \@ifx [1]{%
 \ifx #1\expandafter \@firstoftwo
 \else \expandafter \@secondoftwo
 \fi
}%
\providecommand \natexlab [1]{#1}%
\providecommand \enquote  [1]{``#1''}%
\providecommand \bibnamefont  [1]{#1}%
\providecommand \bibfnamefont [1]{#1}%
\providecommand \citenamefont [1]{#1}%
\providecommand \href@noop [0]{\@secondoftwo}%
\providecommand \href [0]{\begingroup \@sanitize@url \@href}%
\providecommand \@href[1]{\@@startlink{#1}\@@href}%
\providecommand \@@href[1]{\endgroup#1\@@endlink}%
\providecommand \@sanitize@url [0]{\catcode `\\12\catcode `\$12\catcode
  `\&12\catcode `\#12\catcode `\^12\catcode `\_12\catcode `\%12\relax}%
\providecommand \@@startlink[1]{}%
\providecommand \@@endlink[0]{}%
\providecommand \url  [0]{\begingroup\@sanitize@url \@url }%
\providecommand \@url [1]{\endgroup\@href {#1}{\urlprefix }}%
\providecommand \urlprefix  [0]{URL }%
\providecommand \Eprint [0]{\href }%
\providecommand \doibase [0]{http://dx.doi.org/}%
\providecommand \selectlanguage [0]{\@gobble}%
\providecommand \bibinfo  [0]{\@secondoftwo}%
\providecommand \bibfield  [0]{\@secondoftwo}%
\providecommand \translation [1]{[#1]}%
\providecommand \BibitemOpen [0]{}%
\providecommand \bibitemStop [0]{}%
\providecommand \bibitemNoStop [0]{.\EOS\space}%
\providecommand \EOS [0]{\spacefactor3000\relax}%
\providecommand \BibitemShut  [1]{\csname bibitem#1\endcsname}%
\let\auto@bib@innerbib\@empty
\bibitem [{\citenamefont {Arkani-Hamed}\ \emph {et~al.}(2011)\citenamefont
  {Arkani-Hamed}, \citenamefont {Bourjaily}, \citenamefont {Cachazo},
  \citenamefont {Caron-Huot},\ and\ \citenamefont
  {Trnka}}]{ArkaniHamed:2010kv}%
  \BibitemOpen
  \bibfield  {author} {\bibinfo {author} {\bibfnamefont {N.}~\bibnamefont
  {Arkani-Hamed}}, \bibinfo {author} {\bibfnamefont {J.~L.}\ \bibnamefont
  {Bourjaily}}, \bibinfo {author} {\bibfnamefont {F.}~\bibnamefont {Cachazo}},
  \bibinfo {author} {\bibfnamefont {S.}~\bibnamefont {Caron-Huot}}, \ and\
  \bibinfo {author} {\bibfnamefont {J.}~\bibnamefont {Trnka}},\ }\href
  {\doibase 10.1007/JHEP01(2011)041} {\bibfield  {journal} {\bibinfo  {journal}
  {Journal of High Energy Physics}\ }\textbf {\bibinfo {volume} {1101}},\
  \bibinfo {pages} {041} (\bibinfo {year} {2011})},\ \Eprint
  {http://arxiv.org/abs/1008.2958} {arXiv:1008.2958 [hep-th]} \BibitemShut
  {NoStop}%
\bibitem [{\citenamefont {Arkani-Hamed}\ \emph {et~al.}(2010)\citenamefont
  {Arkani-Hamed}, \citenamefont {Cachazo}, \citenamefont {Cheung},\ and\
  \citenamefont {Kaplan}}]{ArkaniHamed:2009dn}%
  \BibitemOpen
  \bibfield  {author} {\bibinfo {author} {\bibfnamefont {N.}~\bibnamefont
  {Arkani-Hamed}}, \bibinfo {author} {\bibfnamefont {F.}~\bibnamefont
  {Cachazo}}, \bibinfo {author} {\bibfnamefont {C.}~\bibnamefont {Cheung}}, \
  and\ \bibinfo {author} {\bibfnamefont {J.}~\bibnamefont {Kaplan}},\ }\href
  {\doibase 10.1007/JHEP03(2010)020} {\bibfield  {journal} {\bibinfo  {journal}
  {JHEP}\ }\textbf {\bibinfo {volume} {03}},\ \bibinfo {pages} {020} (\bibinfo
  {year} {2010})},\ \Eprint {http://arxiv.org/abs/0907.5418} {arXiv:0907.5418
  [hep-th]} \BibitemShut {NoStop}%
\bibitem [{\citenamefont {Arkani-Hamed}\ and\ \citenamefont
  {Trnka}(2014)}]{Arkani-Hamed:2013jha}%
  \BibitemOpen
  \bibfield  {author} {\bibinfo {author} {\bibfnamefont {N.}~\bibnamefont
  {Arkani-Hamed}}\ and\ \bibinfo {author} {\bibfnamefont {J.}~\bibnamefont
  {Trnka}},\ }\href {\doibase 10.1007/JHEP10(2014)030} {\bibfield  {journal}
  {\bibinfo  {journal} {Journal of High Energy Physics}\ }\textbf {\bibinfo
  {volume} {1410}},\ \bibinfo {pages} {30} (\bibinfo {year} {2014})},\ \Eprint
  {http://arxiv.org/abs/1312.2007} {arXiv:1312.2007 [hep-th]} \BibitemShut
  {NoStop}%
\bibitem [{\citenamefont {Arkani-Hamed}\ \emph {et~al.}(2016)\citenamefont
  {Arkani-Hamed}, \citenamefont {Bourjaily}, \citenamefont {Cachazo},
  \citenamefont {Goncharov}, \citenamefont {Postnikov},\ and\ \citenamefont
  {Trnka}}]{Arkani-Hamed:2016byb}%
  \BibitemOpen
  \bibfield  {author} {\bibinfo {author} {\bibfnamefont {N.}~\bibnamefont
  {Arkani-Hamed}}, \bibinfo {author} {\bibfnamefont {J.~L.}\ \bibnamefont
  {Bourjaily}}, \bibinfo {author} {\bibfnamefont {F.}~\bibnamefont {Cachazo}},
  \bibinfo {author} {\bibfnamefont {A.~B.}\ \bibnamefont {Goncharov}}, \bibinfo
  {author} {\bibfnamefont {A.}~\bibnamefont {Postnikov}}, \ and\ \bibinfo
  {author} {\bibfnamefont {J.}~\bibnamefont {Trnka}},\ }\href {\doibase
  10.1017/CBO9781316091548} {\emph {\bibinfo {title} {{Grassmannian Geometry of
  Scattering Amplitudes}}}}\ (\bibinfo  {publisher} {Cambridge University
  Press},\ \bibinfo {year} {2016})\ \Eprint {http://arxiv.org/abs/1212.5605}
  {arXiv:1212.5605 [hep-th]} \BibitemShut {NoStop}%
\bibitem [{\citenamefont {Golden}\ \emph
  {et~al.}(2014{\natexlab{a}})\citenamefont {Golden}, \citenamefont
  {Goncharov}, \citenamefont {Spradlin}, \citenamefont {Vergu},\ and\
  \citenamefont {Volovich}}]{Golden:2013xva}%
  \BibitemOpen
  \bibfield  {author} {\bibinfo {author} {\bibfnamefont {J.}~\bibnamefont
  {Golden}}, \bibinfo {author} {\bibfnamefont {A.~B.}\ \bibnamefont
  {Goncharov}}, \bibinfo {author} {\bibfnamefont {M.}~\bibnamefont {Spradlin}},
  \bibinfo {author} {\bibfnamefont {C.}~\bibnamefont {Vergu}}, \ and\ \bibinfo
  {author} {\bibfnamefont {A.}~\bibnamefont {Volovich}},\ }\href {\doibase
  10.1007/JHEP01(2014)091} {\bibfield  {journal} {\bibinfo  {journal} {JHEP}\
  }\textbf {\bibinfo {volume} {01}},\ \bibinfo {pages} {091} (\bibinfo {year}
  {2014}{\natexlab{a}})},\ \Eprint {http://arxiv.org/abs/1305.1617}
  {arXiv:1305.1617 [hep-th]} \BibitemShut {NoStop}%
\bibitem [{\citenamefont {Dixon}\ \emph {et~al.}(2011)\citenamefont {Dixon},
  \citenamefont {Drummond},\ and\ \citenamefont {Henn}}]{Dixon:2011pw}%
  \BibitemOpen
  \bibfield  {author} {\bibinfo {author} {\bibfnamefont {L.~J.}\ \bibnamefont
  {Dixon}}, \bibinfo {author} {\bibfnamefont {J.~M.}\ \bibnamefont {Drummond}},
  \ and\ \bibinfo {author} {\bibfnamefont {J.~M.}\ \bibnamefont {Henn}},\
  }\href {\doibase 10.1007/JHEP11(2011)023} {\bibfield  {journal} {\bibinfo
  {journal} {Journal of High Energy Physics}\ }\textbf {\bibinfo {volume}
  {2011}},\ \bibinfo {pages} {23} (\bibinfo {year} {2011})},\ \Eprint
  {http://arxiv.org/abs/1108.4461} {arXiv:1108.4461 [hep-th]} \BibitemShut
  {NoStop}%
\bibitem [{\citenamefont {Drummond}\ \emph {et~al.}(2015)\citenamefont
  {Drummond}, \citenamefont {Papathanasiou},\ and\ \citenamefont
  {Spradlin}}]{Drummond:2014ffa}%
  \BibitemOpen
  \bibfield  {author} {\bibinfo {author} {\bibfnamefont {J.~M.}\ \bibnamefont
  {Drummond}}, \bibinfo {author} {\bibfnamefont {G.}~\bibnamefont
  {Papathanasiou}}, \ and\ \bibinfo {author} {\bibfnamefont {M.}~\bibnamefont
  {Spradlin}},\ }\href {\doibase 10.1007/JHEP03(2015)072} {\bibfield  {journal}
  {\bibinfo  {journal} {Journal of High Energy Physics}\ }\textbf {\bibinfo
  {volume} {2015}},\ \bibinfo {pages} {72} (\bibinfo {year} {2015})},\ \Eprint
  {http://arxiv.org/abs/1412.3763} {arXiv:1412.3763 [hep-th]} \BibitemShut
  {NoStop}%
\bibitem [{\citenamefont {Caron-Huot}\ \emph {et~al.}(2016)\citenamefont
  {Caron-Huot}, \citenamefont {Dixon}, \citenamefont {McLeod},\ and\
  \citenamefont {von Hippel}}]{Caron-Huot:2016owq}%
  \BibitemOpen
  \bibfield  {author} {\bibinfo {author} {\bibfnamefont {S.}~\bibnamefont
  {Caron-Huot}}, \bibinfo {author} {\bibfnamefont {L.~J.}\ \bibnamefont
  {Dixon}}, \bibinfo {author} {\bibfnamefont {A.}~\bibnamefont {McLeod}}, \
  and\ \bibinfo {author} {\bibfnamefont {M.}~\bibnamefont {von Hippel}},\
  }\href {\doibase 10.1103/PhysRevLett.117.241601} {\bibfield  {journal}
  {\bibinfo  {journal} {Phys. Rev. Lett.}\ }\textbf {\bibinfo {volume} {117}},\
  \bibinfo {pages} {241601} (\bibinfo {year} {2016})},\ \Eprint
  {http://arxiv.org/abs/1609.00669} {arXiv:1609.00669 [hep-th]} \BibitemShut
  {NoStop}%
\bibitem [{\citenamefont {Caron-Huot}\ \emph {et~al.}(2018)\citenamefont
  {Caron-Huot}, \citenamefont {Dixon}, \citenamefont {von Hippel},
  \citenamefont {McLeod},\ and\ \citenamefont
  {Papathanasiou}}]{Caron-Huot:2018dsv}%
  \BibitemOpen
  \bibfield  {author} {\bibinfo {author} {\bibfnamefont {S.}~\bibnamefont
  {Caron-Huot}}, \bibinfo {author} {\bibfnamefont {L.~J.}\ \bibnamefont
  {Dixon}}, \bibinfo {author} {\bibfnamefont {M.}~\bibnamefont {von Hippel}},
  \bibinfo {author} {\bibfnamefont {A.~J.}\ \bibnamefont {McLeod}}, \ and\
  \bibinfo {author} {\bibfnamefont {G.}~\bibnamefont {Papathanasiou}},\ }\href
  {\doibase 10.1007/JHEP07(2018)170} {\bibfield  {journal} {\bibinfo  {journal}
  {Journal of High Energy Physics}\ }\textbf {\bibinfo {volume} {2018}},\
  \bibinfo {pages} {170} (\bibinfo {year} {2018})},\ \Eprint
  {http://arxiv.org/abs/1806.01361} {arXiv:1806.01361 [hep-th]} \BibitemShut
  {NoStop}%
\bibitem [{\citenamefont {Caron-Huot}\ \emph
  {et~al.}(2019{\natexlab{a}})\citenamefont {Caron-Huot}, \citenamefont
  {Dixon}, \citenamefont {Dulat}, \citenamefont {Von~Hippel}, \citenamefont
  {McLeod},\ and\ \citenamefont {Papathanasiou}}]{Caron-Huot:2019bsq}%
  \BibitemOpen
  \bibfield  {author} {\bibinfo {author} {\bibfnamefont {S.}~\bibnamefont
  {Caron-Huot}}, \bibinfo {author} {\bibfnamefont {L.~J.}\ \bibnamefont
  {Dixon}}, \bibinfo {author} {\bibfnamefont {F.}~\bibnamefont {Dulat}},
  \bibinfo {author} {\bibfnamefont {M.}~\bibnamefont {Von~Hippel}}, \bibinfo
  {author} {\bibfnamefont {A.~J.}\ \bibnamefont {McLeod}}, \ and\ \bibinfo
  {author} {\bibfnamefont {G.}~\bibnamefont {Papathanasiou}},\ }\href {\doibase
  10.1007/JHEP09(2019)061} {\bibfield  {journal} {\bibinfo  {journal} {Journal
  of High Energy Physics}\ }\textbf {\bibinfo {volume} {2019}},\ \bibinfo
  {pages} {61} (\bibinfo {year} {2019}{\natexlab{a}})},\ \Eprint
  {http://arxiv.org/abs/1906.07116} {arXiv:1906.07116 [hep-th]} \BibitemShut
  {NoStop}%
\bibitem [{\citenamefont {Drummond}\ \emph {et~al.}(2018)\citenamefont
  {Drummond}, \citenamefont {Foster},\ and\ \citenamefont
  {G{\"u}rdo{\u{g}}an}}]{Drummond:2017ssj}%
  \BibitemOpen
  \bibfield  {author} {\bibinfo {author} {\bibfnamefont {J.}~\bibnamefont
  {Drummond}}, \bibinfo {author} {\bibfnamefont {J.}~\bibnamefont {Foster}}, \
  and\ \bibinfo {author} {\bibfnamefont {{\"O}.}~\bibnamefont
  {G{\"u}rdo{\u{g}}an}},\ }\href {\doibase 10.1103/PhysRevLett.120.161601}
  {\bibfield  {journal} {\bibinfo  {journal} {Phys. Rev. Lett.}\ }\textbf
  {\bibinfo {volume} {120}},\ \bibinfo {pages} {161601} (\bibinfo {year}
  {2018})},\ \Eprint {http://arxiv.org/abs/1710.10953} {arXiv:1710.10953
  [hep-th]} \BibitemShut {NoStop}%
\bibitem [{\citenamefont {Drummond}\ \emph
  {et~al.}(2019{\natexlab{a}})\citenamefont {Drummond}, \citenamefont
  {Foster},\ and\ \citenamefont {G{\"u}rdo{\u{g}}an}}]{Drummond:2018dfd}%
  \BibitemOpen
  \bibfield  {author} {\bibinfo {author} {\bibfnamefont {J.}~\bibnamefont
  {Drummond}}, \bibinfo {author} {\bibfnamefont {J.}~\bibnamefont {Foster}}, \
  and\ \bibinfo {author} {\bibfnamefont {{\"O}.}~\bibnamefont
  {G{\"u}rdo{\u{g}}an}},\ }\href {\doibase 10.1007/JHEP03(2019)086} {\bibfield
  {journal} {\bibinfo  {journal} {Journal of High Energy Physics}\ }\textbf
  {\bibinfo {volume} {2019}},\ \bibinfo {pages} {86} (\bibinfo {year}
  {2019}{\natexlab{a}})},\ \Eprint {http://arxiv.org/abs/1810.08149}
  {arXiv:1810.08149 [hep-th]} \BibitemShut {NoStop}%
\bibitem [{\citenamefont {Dixon}\ \emph {et~al.}(2017)\citenamefont {Dixon},
  \citenamefont {Drummond}, \citenamefont {Harrington}, \citenamefont {McLeod},
  \citenamefont {Papathanasiou},\ and\ \citenamefont
  {Spradlin}}]{Dixon:2016nkn}%
  \BibitemOpen
  \bibfield  {author} {\bibinfo {author} {\bibfnamefont {L.~J.}\ \bibnamefont
  {Dixon}}, \bibinfo {author} {\bibfnamefont {J.}~\bibnamefont {Drummond}},
  \bibinfo {author} {\bibfnamefont {T.}~\bibnamefont {Harrington}}, \bibinfo
  {author} {\bibfnamefont {A.~J.}\ \bibnamefont {McLeod}}, \bibinfo {author}
  {\bibfnamefont {G.}~\bibnamefont {Papathanasiou}}, \ and\ \bibinfo {author}
  {\bibfnamefont {M.}~\bibnamefont {Spradlin}},\ }\href {\doibase
  10.1007/JHEP02(2017)137} {\bibfield  {journal} {\bibinfo  {journal} {Journal
  of High Energy Physics}\ }\textbf {\bibinfo {volume} {2017}},\ \bibinfo
  {pages} {137} (\bibinfo {year} {2017})},\ \Eprint
  {http://arxiv.org/abs/1612.08976} {arXiv:1612.08976 [hep-th]} \BibitemShut
  {NoStop}%
\bibitem [{\citenamefont {Drummond}\ \emph
  {et~al.}(2019{\natexlab{b}})\citenamefont {Drummond}, \citenamefont {Foster},
  \citenamefont {G{\"u}rdo{\u{g}}an},\ and\ \citenamefont
  {Papathanasiou}}]{Drummond:2018caf}%
  \BibitemOpen
  \bibfield  {author} {\bibinfo {author} {\bibfnamefont {J.}~\bibnamefont
  {Drummond}}, \bibinfo {author} {\bibfnamefont {J.}~\bibnamefont {Foster}},
  \bibinfo {author} {\bibfnamefont {{\"O}.}~\bibnamefont {G{\"u}rdo{\u{g}}an}},
  \ and\ \bibinfo {author} {\bibfnamefont {G.}~\bibnamefont {Papathanasiou}},\
  }\href {\doibase 10.1007/JHEP03(2019)087} {\bibfield  {journal} {\bibinfo
  {journal} {Journal of High Energy Physics}\ }\textbf {\bibinfo {volume}
  {2019}},\ \bibinfo {pages} {87} (\bibinfo {year} {2019}{\natexlab{b}})},\
  \Eprint {http://arxiv.org/abs/1812.04640} {arXiv:1812.04640 [hep-th]}
  \BibitemShut {NoStop}%
\bibitem [{\citenamefont {Caron-Huot}\ \emph
  {et~al.}(2019{\natexlab{b}})\citenamefont {Caron-Huot}, \citenamefont
  {Dixon}, \citenamefont {Dulat}, \citenamefont {von Hippel}, \citenamefont
  {McLeod},\ and\ \citenamefont {Papathanasiou}}]{Caron-Huot:2019vjl}%
  \BibitemOpen
  \bibfield  {author} {\bibinfo {author} {\bibfnamefont {S.}~\bibnamefont
  {Caron-Huot}}, \bibinfo {author} {\bibfnamefont {L.~J.}\ \bibnamefont
  {Dixon}}, \bibinfo {author} {\bibfnamefont {F.}~\bibnamefont {Dulat}},
  \bibinfo {author} {\bibfnamefont {M.}~\bibnamefont {von Hippel}}, \bibinfo
  {author} {\bibfnamefont {A.~J.}\ \bibnamefont {McLeod}}, \ and\ \bibinfo
  {author} {\bibfnamefont {G.}~\bibnamefont {Papathanasiou}},\ }\href {\doibase
  10.1007/JHEP08(2019)016} {\bibfield  {journal} {\bibinfo  {journal} {Journal
  of High Energy Physics}\ }\textbf {\bibinfo {volume} {2019}},\ \bibinfo
  {pages} {16} (\bibinfo {year} {2019}{\natexlab{b}})},\ \Eprint
  {http://arxiv.org/abs/1903.10890} {arXiv:1903.10890 [hep-th]} \BibitemShut
  {NoStop}%
\bibitem [{\citenamefont {Dixon}\ and\ \citenamefont
  {Liu}(2020)}]{Dixon:2020cnr}%
  \BibitemOpen
  \bibfield  {author} {\bibinfo {author} {\bibfnamefont {L.~J.}\ \bibnamefont
  {Dixon}}\ and\ \bibinfo {author} {\bibfnamefont {Y.-T.}\ \bibnamefont
  {Liu}},\ }\href {\doibase 10.1007/JHEP10(2020)031} {\bibfield  {journal}
  {\bibinfo  {journal} {JHEP}\ }\textbf {\bibinfo {volume} {10}},\ \bibinfo
  {pages} {031} (\bibinfo {year} {2020})},\ \Eprint
  {http://arxiv.org/abs/2007.12966} {arXiv:2007.12966 [hep-th]} \BibitemShut
  {NoStop}%
\bibitem [{\citenamefont {Caron-Huot}\ \emph {et~al.}(2020)\citenamefont
  {Caron-Huot}, \citenamefont {Dixon}, \citenamefont {Drummond}, \citenamefont
  {Dulat}, \citenamefont {Foster}, \citenamefont {G\"urdo\u{g}an},
  \citenamefont {von Hippel}, \citenamefont {McLeod},\ and\ \citenamefont
  {Papathanasiou}}]{Caron-Huot:2020bkp}%
  \BibitemOpen
  \bibfield  {author} {\bibinfo {author} {\bibfnamefont {S.}~\bibnamefont
  {Caron-Huot}}, \bibinfo {author} {\bibfnamefont {L.~J.}\ \bibnamefont
  {Dixon}}, \bibinfo {author} {\bibfnamefont {J.~M.}\ \bibnamefont {Drummond}},
  \bibinfo {author} {\bibfnamefont {F.}~\bibnamefont {Dulat}}, \bibinfo
  {author} {\bibfnamefont {J.}~\bibnamefont {Foster}}, \bibinfo {author}
  {\bibfnamefont {O.}~\bibnamefont {G\"urdo\u{g}an}}, \bibinfo {author}
  {\bibfnamefont {M.}~\bibnamefont {von Hippel}}, \bibinfo {author}
  {\bibfnamefont {A.~J.}\ \bibnamefont {McLeod}}, \ and\ \bibinfo {author}
  {\bibfnamefont {G.}~\bibnamefont {Papathanasiou}},\ }\href {\doibase
  10.22323/1.376.0003} {\bibfield  {journal} {\bibinfo  {journal} {PoS}\
  }\textbf {\bibinfo {volume} {CORFU2019}},\ \bibinfo {pages} {003} (\bibinfo
  {year} {2020})},\ \Eprint {http://arxiv.org/abs/2005.06735} {arXiv:2005.06735
  [hep-th]} \BibitemShut {NoStop}%
\bibitem [{\citenamefont {Drummond}\ \emph {et~al.}(2010)\citenamefont
  {Drummond}, \citenamefont {Henn}, \citenamefont {Korchemsky},\ and\
  \citenamefont {Sokatchev}}]{Drummond:2008vq}%
  \BibitemOpen
  \bibfield  {author} {\bibinfo {author} {\bibfnamefont {J.}~\bibnamefont
  {Drummond}}, \bibinfo {author} {\bibfnamefont {J.}~\bibnamefont {Henn}},
  \bibinfo {author} {\bibfnamefont {G.}~\bibnamefont {Korchemsky}}, \ and\
  \bibinfo {author} {\bibfnamefont {E.}~\bibnamefont {Sokatchev}},\ }\href
  {\doibase 10.1016/j.nuclphysb.2009.11.022} {\bibfield  {journal} {\bibinfo
  {journal} {Nucl.Phys.}\ }\textbf {\bibinfo {volume} {B828}},\ \bibinfo
  {pages} {317} (\bibinfo {year} {2010})},\ \Eprint
  {http://arxiv.org/abs/0807.1095} {arXiv:0807.1095 [hep-th]} \BibitemShut
  {NoStop}%
\bibitem [{\citenamefont {Berkovits}\ and\ \citenamefont
  {Maldacena}(2008)}]{Berkovits:2008ic}%
  \BibitemOpen
  \bibfield  {author} {\bibinfo {author} {\bibfnamefont {N.}~\bibnamefont
  {Berkovits}}\ and\ \bibinfo {author} {\bibfnamefont {J.}~\bibnamefont
  {Maldacena}},\ }\href {\doibase 10.1088/1126-6708/2008/09/062} {\bibfield
  {journal} {\bibinfo  {journal} {Journal of High Energy Physics}\ }\textbf
  {\bibinfo {volume} {0809}},\ \bibinfo {pages} {062} (\bibinfo {year}
  {2008})},\ \Eprint {http://arxiv.org/abs/0807.3196} {arXiv:0807.3196
  [hep-th]} \BibitemShut {NoStop}%
\bibitem [{\citenamefont {Drummond}\ \emph {et~al.}(2009)\citenamefont
  {Drummond}, \citenamefont {Henn},\ and\ \citenamefont
  {Plefka}}]{Drummond:2009fd}%
  \BibitemOpen
  \bibfield  {author} {\bibinfo {author} {\bibfnamefont {J.~M.}\ \bibnamefont
  {Drummond}}, \bibinfo {author} {\bibfnamefont {J.~M.}\ \bibnamefont {Henn}},
  \ and\ \bibinfo {author} {\bibfnamefont {J.}~\bibnamefont {Plefka}},\ }\href
  {\doibase 10.1088/1126-6708/2009/05/046} {\bibfield  {journal} {\bibinfo
  {journal} {Journal of High Energy Physics}\ }\textbf {\bibinfo {volume}
  {0905}},\ \bibinfo {pages} {046} (\bibinfo {year} {2009})},\ \Eprint
  {http://arxiv.org/abs/0902.2987} {arXiv:0902.2987 [hep-th]} \BibitemShut
  {NoStop}%
\bibitem [{\citenamefont {Arkani-Hamed}\ \emph {et~al.}(2012)\citenamefont
  {Arkani-Hamed}, \citenamefont {Bourjaily}, \citenamefont {Cachazo},\ and\
  \citenamefont {Trnka}}]{ArkaniHamed:2010gh}%
  \BibitemOpen
  \bibfield  {author} {\bibinfo {author} {\bibfnamefont {N.}~\bibnamefont
  {Arkani-Hamed}}, \bibinfo {author} {\bibfnamefont {J.~L.}\ \bibnamefont
  {Bourjaily}}, \bibinfo {author} {\bibfnamefont {F.}~\bibnamefont {Cachazo}},
  \ and\ \bibinfo {author} {\bibfnamefont {J.}~\bibnamefont {Trnka}},\ }\href
  {\doibase 10.1007/JHEP06(2012)125} {\bibfield  {journal} {\bibinfo  {journal}
  {JHEP}\ }\textbf {\bibinfo {volume} {06}},\ \bibinfo {pages} {125} (\bibinfo
  {year} {2012})},\ \Eprint {http://arxiv.org/abs/1012.6032} {arXiv:1012.6032
  [hep-th]} \BibitemShut {NoStop}%
\bibitem [{\citenamefont {Henn}(2013)}]{Henn:2013pwa}%
  \BibitemOpen
  \bibfield  {author} {\bibinfo {author} {\bibfnamefont {J.~M.}\ \bibnamefont
  {Henn}},\ }\href {\doibase 10.1103/PhysRevLett.110.251601} {\bibfield
  {journal} {\bibinfo  {journal} {Phys. Rev. Lett.}\ }\textbf {\bibinfo
  {volume} {110}},\ \bibinfo {pages} {251601} (\bibinfo {year} {2013})},\
  \Eprint {http://arxiv.org/abs/1304.1806} {arXiv:1304.1806 [hep-th]}
  \BibitemShut {NoStop}%
\bibitem [{\citenamefont {Goncharov}\ \emph {et~al.}(2010)\citenamefont
  {Goncharov}, \citenamefont {Spradlin}, \citenamefont {Vergu},\ and\
  \citenamefont {Volovich}}]{Goncharov:2010jf}%
  \BibitemOpen
  \bibfield  {author} {\bibinfo {author} {\bibfnamefont {A.~B.}\ \bibnamefont
  {Goncharov}}, \bibinfo {author} {\bibfnamefont {M.}~\bibnamefont {Spradlin}},
  \bibinfo {author} {\bibfnamefont {C.}~\bibnamefont {Vergu}}, \ and\ \bibinfo
  {author} {\bibfnamefont {A.}~\bibnamefont {Volovich}},\ }\href {\doibase
  10.1103/PhysRevLett.105.151605} {\bibfield  {journal} {\bibinfo  {journal}
  {Phys.Rev.Lett.}\ }\textbf {\bibinfo {volume} {105}},\ \bibinfo {pages}
  {151605} (\bibinfo {year} {2010})},\ \Eprint {http://arxiv.org/abs/1006.5703}
  {arXiv:1006.5703 [hep-th]} \BibitemShut {NoStop}%
\bibitem [{\citenamefont {Duhr}(2019)}]{Duhr:2019wtr}%
  \BibitemOpen
  \bibfield  {author} {\bibinfo {author} {\bibfnamefont {C.}~\bibnamefont
  {Duhr}},\ }\href {\doibase 10.1146/annurev-nucl-101918-023551} {\bibfield
  {journal} {\bibinfo  {journal} {Ann. Rev. Nucl. Part. Sci.}\ }\textbf
  {\bibinfo {volume} {69}},\ \bibinfo {pages} {15} (\bibinfo {year}
  {2019})}\BibitemShut {NoStop}%
\bibitem [{\citenamefont {Henn}(2020)}]{Henn:2020omi}%
  \BibitemOpen
  \bibfield  {author} {\bibinfo {author} {\bibfnamefont {J.~M.}\ \bibnamefont
  {Henn}},\ }\href@noop {} {\  (\bibinfo {year} {2020})},\ \Eprint
  {http://arxiv.org/abs/2006.00361} {arXiv:2006.00361 [hep-th]} \BibitemShut
  {NoStop}%
\bibitem [{\citenamefont {Fomin}\ and\ \citenamefont
  {Zelevinsky}(2002)}]{1021.16017}%
  \BibitemOpen
  \bibfield  {author} {\bibinfo {author} {\bibfnamefont {S.}~\bibnamefont
  {Fomin}}\ and\ \bibinfo {author} {\bibfnamefont {A.}~\bibnamefont
  {Zelevinsky}},\ }\href {\doibase 10.1090/S0894-0347-01-00385-X} {\bibfield
  {journal} {\bibinfo  {journal} {Journal of the American Mathematical
  Society}\ }\textbf {\bibinfo {volume} {15}},\ \bibinfo {pages} {497}
  (\bibinfo {year} {2002})},\ \Eprint {http://arxiv.org/abs/math/0104151}
  {arXiv:math/0104151 [math.RT]} \BibitemShut {NoStop}%
\bibitem [{\citenamefont {Fomin}\ and\ \citenamefont
  {Zelevinsky}(2003)}]{1054.17024}%
  \BibitemOpen
  \bibfield  {author} {\bibinfo {author} {\bibfnamefont {S.}~\bibnamefont
  {Fomin}}\ and\ \bibinfo {author} {\bibfnamefont {A.}~\bibnamefont
  {Zelevinsky}},\ }\href {\doibase 10.1007/s00222-003-0302-y} {\bibfield
  {journal} {\bibinfo  {journal} {Inventiones mathematicae}\ }\textbf {\bibinfo
  {volume} {154}},\ \bibinfo {pages} {63} (\bibinfo {year} {2003})},\ \Eprint
  {http://arxiv.org/abs/math/0208229} {arXiv:math/0208229 [math.RA]}
  \BibitemShut {NoStop}%
\bibitem [{\citenamefont {Berenstein}\ \emph {et~al.}(2003)\citenamefont
  {Berenstein}, \citenamefont {Fomin},\ and\ \citenamefont
  {Zelevinsky}}]{CAIII}%
  \BibitemOpen
  \bibfield  {author} {\bibinfo {author} {\bibfnamefont {A.}~\bibnamefont
  {Berenstein}}, \bibinfo {author} {\bibfnamefont {S.}~\bibnamefont {Fomin}}, \
  and\ \bibinfo {author} {\bibfnamefont {A.}~\bibnamefont {Zelevinsky}},\
  }\href {\doibase 10.1215/S0012-7094-04-12611-9} {\bibfield  {journal}
  {\bibinfo  {journal} {Duke Mathematical Journal}\ }\textbf {\bibinfo {volume}
  {126}},\ \bibinfo {pages} {1} (\bibinfo {year} {2003})},\ \Eprint
  {http://arxiv.org/abs/math/0305434} {arXiv:math/0305434 [math.RT]}
  \BibitemShut {NoStop}%
\bibitem [{\citenamefont {Fomin}\ and\ \citenamefont
  {Zelevinsky}(2007)}]{CAIV}%
  \BibitemOpen
  \bibfield  {author} {\bibinfo {author} {\bibfnamefont {S.}~\bibnamefont
  {Fomin}}\ and\ \bibinfo {author} {\bibfnamefont {A.}~\bibnamefont
  {Zelevinsky}},\ }\href {\doibase 10.1112/S0010437X06002521} {\bibfield
  {journal} {\bibinfo  {journal} {Compositio Mathematica}\ }\textbf {\bibinfo
  {volume} {143}},\ \bibinfo {pages} {112} (\bibinfo {year} {2007})},\ \Eprint
  {http://arxiv.org/abs/math/0602259} {arXiv:math/0602259 [math.RA]}
  \BibitemShut {NoStop}%
\bibitem [{\citenamefont {{Keller}}(2008)}]{2008arXiv0807.1960K}%
  \BibitemOpen
  \bibfield  {author} {\bibinfo {author} {\bibfnamefont {B.}~\bibnamefont
  {{Keller}}},\ }\href@noop {} {\bibfield  {journal} {\bibinfo  {journal}
  {arXiv e-prints}\ ,\ \bibinfo {eid} {arXiv:0807.1960}} (\bibinfo {year}
  {2008})},\ \Eprint {http://arxiv.org/abs/0807.1960} {arXiv:0807.1960
  [math.RT]} \BibitemShut {NoStop}%
\bibitem [{\citenamefont {Lampe}()}]{Lampe13}%
  \BibitemOpen
  \bibfield  {author} {\bibinfo {author} {\bibfnamefont {P.}~\bibnamefont
  {Lampe}},\ }\href@noop {} {\enquote {\bibinfo {title} {{Cluster Algebras}},}\
  }\bibinfo {howpublished} {4 December 2013,
  \href{http://www.math.uni-bielefeld.de/~lampe/teaching/cluster/cluster.pdf}{http://www.math.uni-bielefeld.de/~lampe/teaching/cluster/cluster.pdf}}\BibitemShut
  {NoStop}%
\bibitem [{\citenamefont {Fomin}\ \emph {et~al.}(2016)\citenamefont {Fomin},
  \citenamefont {Williams},\ and\ \citenamefont {Zelevinsky}}]{Fomin2016}%
  \BibitemOpen
  \bibfield  {author} {\bibinfo {author} {\bibfnamefont {S.}~\bibnamefont
  {Fomin}}, \bibinfo {author} {\bibfnamefont {L.}~\bibnamefont {Williams}}, \
  and\ \bibinfo {author} {\bibfnamefont {A.}~\bibnamefont {Zelevinsky}},\
  }\href@noop {} {\  (\bibinfo {year} {2016})},\ \Eprint
  {http://arxiv.org/abs/1608.05735} {arXiv:1608.05735 [math.CO]} \BibitemShut
  {NoStop}%
\bibitem [{\citenamefont {Fomin}\ \emph {et~al.}(2017)\citenamefont {Fomin},
  \citenamefont {Williams},\ and\ \citenamefont {Zelevinsky}}]{Fomin2017}%
  \BibitemOpen
  \bibfield  {author} {\bibinfo {author} {\bibfnamefont {S.}~\bibnamefont
  {Fomin}}, \bibinfo {author} {\bibfnamefont {L.}~\bibnamefont {Williams}}, \
  and\ \bibinfo {author} {\bibfnamefont {A.}~\bibnamefont {Zelevinsky}},\
  }\href@noop {} {\  (\bibinfo {year} {2017})},\ \Eprint
  {http://arxiv.org/abs/1707.07190} {arXiv:1707.07190 [math.CO]} \BibitemShut
  {NoStop}%
\bibitem [{\citenamefont {Fock}\ and\ \citenamefont {Goncharov}(2009)}]{FG03b}%
  \BibitemOpen
  \bibfield  {author} {\bibinfo {author} {\bibfnamefont {V.~V.}\ \bibnamefont
  {Fock}}\ and\ \bibinfo {author} {\bibfnamefont {A.~B.}\ \bibnamefont
  {Goncharov}},\ }\href@noop {} {\bibfield  {journal} {\bibinfo  {journal}
  {Ann. Sci. \'Ec. Norm. Sup\'er. (4)}\ }\textbf {\bibinfo {volume} {42}},\
  \bibinfo {pages} {865} (\bibinfo {year} {2009})},\ \Eprint
  {http://arxiv.org/abs/math/0311245} {arXiv:math/0311245 [math.AG]}
  \BibitemShut {NoStop}%
\bibitem [{\citenamefont {Parker}\ \emph {et~al.}(2015)\citenamefont {Parker},
  \citenamefont {Scherlis}, \citenamefont {Spradlin},\ and\ \citenamefont
  {Volovich}}]{Parker:2015cia}%
  \BibitemOpen
  \bibfield  {author} {\bibinfo {author} {\bibfnamefont {D.}~\bibnamefont
  {Parker}}, \bibinfo {author} {\bibfnamefont {A.}~\bibnamefont {Scherlis}},
  \bibinfo {author} {\bibfnamefont {M.}~\bibnamefont {Spradlin}}, \ and\
  \bibinfo {author} {\bibfnamefont {A.}~\bibnamefont {Volovich}},\ }\href
  {\doibase 10.1007/JHEP11(2015)136} {\bibfield  {journal} {\bibinfo  {journal}
  {Journal of High Energy Physics}\ }\textbf {\bibinfo {volume} {11}},\
  \bibinfo {pages} {136} (\bibinfo {year} {2015})},\ \Eprint
  {http://arxiv.org/abs/1507.01950} {arXiv:1507.01950 [hep-th]} \BibitemShut
  {NoStop}%
\bibitem [{\citenamefont {Chen}(1977)}]{chen1977}%
  \BibitemOpen
  \bibfield  {author} {\bibinfo {author} {\bibfnamefont {K.-T.}\ \bibnamefont
  {Chen}},\ }\href {\doibase 10.1090/S0002-9904-1977-14320-6} {\bibfield
  {journal} {\bibinfo  {journal} {Bull. Amer. Math. Soc.}\ }\textbf {\bibinfo
  {volume} {83}},\ \bibinfo {pages} {831} (\bibinfo {year} {1977})}\BibitemShut
  {NoStop}%
\bibitem [{\citenamefont {Gehrmann}\ \emph {et~al.}(2012)\citenamefont
  {Gehrmann}, \citenamefont {Jaquier}, \citenamefont {Glover},\ and\
  \citenamefont {Koukoutsakis}}]{Gehrmann:2011aa}%
  \BibitemOpen
  \bibfield  {author} {\bibinfo {author} {\bibfnamefont {T.}~\bibnamefont
  {Gehrmann}}, \bibinfo {author} {\bibfnamefont {M.}~\bibnamefont {Jaquier}},
  \bibinfo {author} {\bibfnamefont {E.}~\bibnamefont {Glover}}, \ and\ \bibinfo
  {author} {\bibfnamefont {A.}~\bibnamefont {Koukoutsakis}},\ }\href {\doibase
  10.1007/JHEP02(2012)056} {\bibfield  {journal} {\bibinfo  {journal} {JHEP}\
  }\textbf {\bibinfo {volume} {02}},\ \bibinfo {pages} {056} (\bibinfo {year}
  {2012})},\ \Eprint {http://arxiv.org/abs/1112.3554} {arXiv:1112.3554
  [hep-ph]} \BibitemShut {NoStop}%
\bibitem [{\citenamefont {Duhr}(2012)}]{Duhr:2012fh}%
  \BibitemOpen
  \bibfield  {author} {\bibinfo {author} {\bibfnamefont {C.}~\bibnamefont
  {Duhr}},\ }\href {\doibase 10.1007/JHEP08(2012)043} {\bibfield  {journal}
  {\bibinfo  {journal} {JHEP}\ }\textbf {\bibinfo {volume} {08}},\ \bibinfo
  {pages} {043} (\bibinfo {year} {2012})},\ \Eprint
  {http://arxiv.org/abs/1203.0454} {arXiv:1203.0454 [hep-ph]} \BibitemShut
  {NoStop}%
\bibitem [{\citenamefont {Gehrmann}\ \emph {et~al.}(2013)\citenamefont
  {Gehrmann}, \citenamefont {Tancredi},\ and\ \citenamefont
  {Weihs}}]{Gehrmann:2013vga}%
  \BibitemOpen
  \bibfield  {author} {\bibinfo {author} {\bibfnamefont {T.}~\bibnamefont
  {Gehrmann}}, \bibinfo {author} {\bibfnamefont {L.}~\bibnamefont {Tancredi}},
  \ and\ \bibinfo {author} {\bibfnamefont {E.}~\bibnamefont {Weihs}},\ }\href
  {\doibase 10.1007/JHEP04(2013)101} {\bibfield  {journal} {\bibinfo  {journal}
  {JHEP}\ }\textbf {\bibinfo {volume} {04}},\ \bibinfo {pages} {101} (\bibinfo
  {year} {2013})},\ \Eprint {http://arxiv.org/abs/1302.2630} {arXiv:1302.2630
  [hep-ph]} \BibitemShut {NoStop}%
\bibitem [{\citenamefont {Brandhuber}\ \emph {et~al.}(2017)\citenamefont
  {Brandhuber}, \citenamefont {Kostacinska}, \citenamefont {Penante},\ and\
  \citenamefont {Travaglini}}]{Brandhuber:2017bkg}%
  \BibitemOpen
  \bibfield  {author} {\bibinfo {author} {\bibfnamefont {A.}~\bibnamefont
  {Brandhuber}}, \bibinfo {author} {\bibfnamefont {M.}~\bibnamefont
  {Kostacinska}}, \bibinfo {author} {\bibfnamefont {B.}~\bibnamefont
  {Penante}}, \ and\ \bibinfo {author} {\bibfnamefont {G.}~\bibnamefont
  {Travaglini}},\ }\href {\doibase 10.1103/PhysRevLett.119.161601} {\bibfield
  {journal} {\bibinfo  {journal} {Phys. Rev. Lett.}\ }\textbf {\bibinfo
  {volume} {119}},\ \bibinfo {pages} {161601} (\bibinfo {year} {2017})},\
  \Eprint {http://arxiv.org/abs/1707.09897} {arXiv:1707.09897 [hep-th]}
  \BibitemShut {NoStop}%
\bibitem [{\citenamefont {Sever}\ \emph {et~al.}(2020)\citenamefont {Sever},
  \citenamefont {Tumanov},\ and\ \citenamefont {Wilhelm}}]{Sever:2020jjx}%
  \BibitemOpen
  \bibfield  {author} {\bibinfo {author} {\bibfnamefont {A.}~\bibnamefont
  {Sever}}, \bibinfo {author} {\bibfnamefont {A.~G.}\ \bibnamefont {Tumanov}},
  \ and\ \bibinfo {author} {\bibfnamefont {M.}~\bibnamefont {Wilhelm}},\
  }\href@noop {} {\  (\bibinfo {year} {2020})},\ \Eprint
  {http://arxiv.org/abs/2009.11297} {arXiv:2009.11297 [hep-th]} \BibitemShut
  {NoStop}%
\bibitem [{\citenamefont {Gehrmann}\ and\ \citenamefont
  {Remiddi}(2001{\natexlab{a}})}]{Gehrmann:2000zt}%
  \BibitemOpen
  \bibfield  {author} {\bibinfo {author} {\bibfnamefont {T.}~\bibnamefont
  {Gehrmann}}\ and\ \bibinfo {author} {\bibfnamefont {E.}~\bibnamefont
  {Remiddi}},\ }\href {\doibase 10.1016/S0550-3213(01)00057-8} {\bibfield
  {journal} {\bibinfo  {journal} {Nucl. Phys. B}\ }\textbf {\bibinfo {volume}
  {601}},\ \bibinfo {pages} {248} (\bibinfo {year} {2001}{\natexlab{a}})},\
  \Eprint {http://arxiv.org/abs/hep-ph/0008287} {arXiv:hep-ph/0008287}
  \BibitemShut {NoStop}%
\bibitem [{\citenamefont {Gehrmann}\ and\ \citenamefont
  {Remiddi}(2001{\natexlab{b}})}]{Gehrmann:2001ck}%
  \BibitemOpen
  \bibfield  {author} {\bibinfo {author} {\bibfnamefont {T.}~\bibnamefont
  {Gehrmann}}\ and\ \bibinfo {author} {\bibfnamefont {E.}~\bibnamefont
  {Remiddi}},\ }\href {\doibase 10.1016/S0550-3213(01)00074-8} {\bibfield
  {journal} {\bibinfo  {journal} {Nucl. Phys. B}\ }\textbf {\bibinfo {volume}
  {601}},\ \bibinfo {pages} {287} (\bibinfo {year} {2001}{\natexlab{b}})},\
  \Eprint {http://arxiv.org/abs/hep-ph/0101124} {arXiv:hep-ph/0101124}
  \BibitemShut {NoStop}%
\bibitem [{\citenamefont {Di~Vita}\ \emph {et~al.}(2014)\citenamefont
  {Di~Vita}, \citenamefont {Mastrolia}, \citenamefont {Schubert},\ and\
  \citenamefont {Yundin}}]{DiVita:2014pza}%
  \BibitemOpen
  \bibfield  {author} {\bibinfo {author} {\bibfnamefont {S.}~\bibnamefont
  {Di~Vita}}, \bibinfo {author} {\bibfnamefont {P.}~\bibnamefont {Mastrolia}},
  \bibinfo {author} {\bibfnamefont {U.}~\bibnamefont {Schubert}}, \ and\
  \bibinfo {author} {\bibfnamefont {V.}~\bibnamefont {Yundin}},\ }\href
  {\doibase 10.1007/JHEP09(2014)148} {\bibfield  {journal} {\bibinfo  {journal}
  {JHEP}\ }\textbf {\bibinfo {volume} {09}},\ \bibinfo {pages} {148} (\bibinfo
  {year} {2014})},\ \Eprint {http://arxiv.org/abs/1408.3107} {arXiv:1408.3107
  [hep-ph]} \BibitemShut {NoStop}%
\bibitem [{\citenamefont {Gehrmann}\ and\ \citenamefont
  {Remiddi}(2002)}]{Gehrmann:2001jv}%
  \BibitemOpen
  \bibfield  {author} {\bibinfo {author} {\bibfnamefont {T.}~\bibnamefont
  {Gehrmann}}\ and\ \bibinfo {author} {\bibfnamefont {E.}~\bibnamefont
  {Remiddi}},\ }\href {\doibase 10.1016/S0010-4655(02)00139-X} {\bibfield
  {journal} {\bibinfo  {journal} {Comput. Phys. Commun.}\ }\textbf {\bibinfo
  {volume} {144}},\ \bibinfo {pages} {200} (\bibinfo {year} {2002})},\ \Eprint
  {http://arxiv.org/abs/hep-ph/0111255} {arXiv:hep-ph/0111255} \BibitemShut
  {NoStop}%
\bibitem [{Note1()}]{Note1}%
  \BibitemOpen
  \bibinfo {note} {At one loop, the five-letter subalphabet $\protect
  \{z_1,z_2,1-z_1,1-z_2,z_1+z_2\protect \}$ is sufficient. It is well-known
  that it corresponds to the $A_{2}$ cluster algebra. Closely related function
  spaces have also appeared in off-shell form factors, four-point CFT
  correlation functions, and in the soft anomalous dimension
  matrix.}\BibitemShut {Stop}%
\bibitem [{\citenamefont {Golden}\ \emph
  {et~al.}(2014{\natexlab{b}})\citenamefont {Golden}, \citenamefont {Paulos},
  \citenamefont {Spradlin},\ and\ \citenamefont {Volovich}}]{Golden:2014xqa}%
  \BibitemOpen
  \bibfield  {author} {\bibinfo {author} {\bibfnamefont {J.}~\bibnamefont
  {Golden}}, \bibinfo {author} {\bibfnamefont {M.~F.}\ \bibnamefont {Paulos}},
  \bibinfo {author} {\bibfnamefont {M.}~\bibnamefont {Spradlin}}, \ and\
  \bibinfo {author} {\bibfnamefont {A.}~\bibnamefont {Volovich}},\ }\href
  {\doibase 10.1088/1751-8113/47/47/474005} {\bibfield  {journal} {\bibinfo
  {journal} {J. Phys.}\ }\textbf {\bibinfo {volume} {A47}},\ \bibinfo {pages}
  {474005} (\bibinfo {year} {2014}{\natexlab{b}})},\ \Eprint
  {http://arxiv.org/abs/1401.6446} {arXiv:1401.6446 [hep-th]} \BibitemShut
  {NoStop}%
\bibitem [{\citenamefont {Dixon}\ \emph {et~al.}(2020)\citenamefont {Dixon},
  \citenamefont {McLeod},\ and\ \citenamefont {Wilhelm}}]{Dixon:2020bbt}%
  \BibitemOpen
  \bibfield  {author} {\bibinfo {author} {\bibfnamefont {L.~J.}\ \bibnamefont
  {Dixon}}, \bibinfo {author} {\bibfnamefont {A.~J.}\ \bibnamefont {McLeod}}, \
  and\ \bibinfo {author} {\bibfnamefont {M.}~\bibnamefont {Wilhelm}},\
  }\href@noop {} {\  (\bibinfo {year} {2020})},\ \Eprint
  {http://arxiv.org/abs/2012.12286} {arXiv:2012.12286 [hep-th]} \BibitemShut
  {NoStop}%
\bibitem [{\citenamefont {Jin}\ and\ \citenamefont {Yang}(2018)}]{Jin:2018fak}%
  \BibitemOpen
  \bibfield  {author} {\bibinfo {author} {\bibfnamefont {Q.}~\bibnamefont
  {Jin}}\ and\ \bibinfo {author} {\bibfnamefont {G.}~\bibnamefont {Yang}},\
  }\href {\doibase 10.1103/PhysRevLett.121.101603} {\bibfield  {journal}
  {\bibinfo  {journal} {Phys. Rev. Lett.}\ }\textbf {\bibinfo {volume} {121}},\
  \bibinfo {pages} {101603} (\bibinfo {year} {2018})},\ \Eprint
  {http://arxiv.org/abs/1804.04653} {arXiv:1804.04653 [hep-th]} \BibitemShut
  {NoStop}%
\bibitem [{Note2()}]{Note2}%
  \BibitemOpen
  \bibinfo {note} {More generally, we find that $C_{n}$ can be identified as
  the parity-invariant surface inside $A_{2n-1}$, and that the same holds true
  for $F_4$ inside $E_{6}$. Similarly, we have obtained $G_{2}$ and $B_{n}$
  from $D_{4}$ and $D_{2n-1}$, respectively. See also \cite
  {Arkani-Hamed:2020tuz}.}\BibitemShut {Stop}%
\bibitem [{Note3()}]{Note3}%
  \BibitemOpen
  \bibinfo {note} {Note however that the first entries are $z_{i}$ with
  $i=1,2,3$, while the adjacency conditions apply to the $1-z_{i}$ entries,
  i.e. the relations found cannot be interpreted as Steinmann relations in the
  kinematic space of four-particle scattering with one off-shell
  leg.}\BibitemShut {Stop}%
\bibitem [{Note4()}]{Note4}%
  \BibitemOpen
  \bibinfo {note} {It is very interesting to note that the $A_{3}$ mutations
  that lie on the parity-invariant surface relate $a_{2i-1}\leftrightarrow
  a_{2i+1}$, i.e. exactly the pairs appearing in the adjacency
  relations~\protect \textup {\hbox {\mathsurround \z@ \protect \normalfont
  (\ignorespaces \ref {eq:ObservedAdjacency}\unskip \@@italiccorr )}} we
  observe.}\BibitemShut {Stop}%
\bibitem [{\citenamefont {Henn}\ \emph {et~al.}(2013)\citenamefont {Henn},
  \citenamefont {Smirnov},\ and\ \citenamefont {Smirnov}}]{Henn:2013fah}%
  \BibitemOpen
  \bibfield  {author} {\bibinfo {author} {\bibfnamefont {J.~M.}\ \bibnamefont
  {Henn}}, \bibinfo {author} {\bibfnamefont {A.~V.}\ \bibnamefont {Smirnov}}, \
  and\ \bibinfo {author} {\bibfnamefont {V.~A.}\ \bibnamefont {Smirnov}},\
  }\href {\doibase 10.1007/JHEP07(2013)128} {\bibfield  {journal} {\bibinfo
  {journal} {JHEP}\ }\textbf {\bibinfo {volume} {07}},\ \bibinfo {pages} {128}
  (\bibinfo {year} {2013})},\ \Eprint {http://arxiv.org/abs/1306.2799}
  {arXiv:1306.2799 [hep-th]} \BibitemShut {NoStop}%
\bibitem [{\citenamefont {Henn}\ \emph {et~al.}(2014)\citenamefont {Henn},
  \citenamefont {Smirnov},\ and\ \citenamefont {Smirnov}}]{Henn:2013nsa}%
  \BibitemOpen
  \bibfield  {author} {\bibinfo {author} {\bibfnamefont {J.~M.}\ \bibnamefont
  {Henn}}, \bibinfo {author} {\bibfnamefont {A.~V.}\ \bibnamefont {Smirnov}}, \
  and\ \bibinfo {author} {\bibfnamefont {V.~A.}\ \bibnamefont {Smirnov}},\
  }\href {\doibase 10.1007/JHEP03(2014)088} {\bibfield  {journal} {\bibinfo
  {journal} {JHEP}\ }\textbf {\bibinfo {volume} {03}},\ \bibinfo {pages} {088}
  (\bibinfo {year} {2014})},\ \Eprint {http://arxiv.org/abs/1312.2588}
  {arXiv:1312.2588 [hep-th]} \BibitemShut {NoStop}%
\bibitem [{\citenamefont {Henn}\ \emph {et~al.}(2020)\citenamefont {Henn},
  \citenamefont {Mistlberger}, \citenamefont {Smirnov},\ and\ \citenamefont
  {Wasser}}]{Henn:2020lye}%
  \BibitemOpen
  \bibfield  {author} {\bibinfo {author} {\bibfnamefont {J.}~\bibnamefont
  {Henn}}, \bibinfo {author} {\bibfnamefont {B.}~\bibnamefont {Mistlberger}},
  \bibinfo {author} {\bibfnamefont {V.~A.}\ \bibnamefont {Smirnov}}, \ and\
  \bibinfo {author} {\bibfnamefont {P.}~\bibnamefont {Wasser}},\ }\href
  {\doibase 10.1007/JHEP04(2020)167} {\bibfield  {journal} {\bibinfo  {journal}
  {JHEP}\ }\textbf {\bibinfo {volume} {04}},\ \bibinfo {pages} {167} (\bibinfo
  {year} {2020})},\ \Eprint {http://arxiv.org/abs/2002.09492} {arXiv:2002.09492
  [hep-ph]} \BibitemShut {NoStop}%
\bibitem [{\citenamefont {Henn}\ and\ \citenamefont
  {Smirnov}(2013)}]{Henn:2013woa}%
  \BibitemOpen
  \bibfield  {author} {\bibinfo {author} {\bibfnamefont {J.~M.}\ \bibnamefont
  {Henn}}\ and\ \bibinfo {author} {\bibfnamefont {V.~A.}\ \bibnamefont
  {Smirnov}},\ }\href {\doibase 10.1007/JHEP11(2013)041} {\bibfield  {journal}
  {\bibinfo  {journal} {JHEP}\ }\textbf {\bibinfo {volume} {11}},\ \bibinfo
  {pages} {041} (\bibinfo {year} {2013})},\ \Eprint
  {http://arxiv.org/abs/1307.4083} {arXiv:1307.4083 [hep-th]} \BibitemShut
  {NoStop}%
\bibitem [{Note5()}]{Note5}%
  \BibitemOpen
  \bibinfo {note} {In fact, only 8 independent combinations of the letters
  appear in the one-loop differential equations.}\BibitemShut {Stop}%
\bibitem [{\citenamefont {Remiddi}\ and\ \citenamefont
  {Vermaseren}(2000)}]{Remiddi:1999ew}%
  \BibitemOpen
  \bibfield  {author} {\bibinfo {author} {\bibfnamefont {E.}~\bibnamefont
  {Remiddi}}\ and\ \bibinfo {author} {\bibfnamefont {J.}~\bibnamefont
  {Vermaseren}},\ }\href {\doibase 10.1142/S0217751X00000367} {\bibfield
  {journal} {\bibinfo  {journal} {Int.J.Mod.Phys.}\ }\textbf {\bibinfo {volume}
  {A15}},\ \bibinfo {pages} {725} (\bibinfo {year} {2000})},\ \Eprint
  {http://arxiv.org/abs/hep-ph/9905237} {arXiv:hep-ph/9905237 [hep-ph]}
  \BibitemShut {NoStop}%
\bibitem [{\citenamefont {Del~Duca}\ \emph
  {et~al.}(2011{\natexlab{a}})\citenamefont {Del~Duca}, \citenamefont {Duhr},\
  and\ \citenamefont {Smirnov}}]{DelDuca:2011jm}%
  \BibitemOpen
  \bibfield  {author} {\bibinfo {author} {\bibfnamefont {V.}~\bibnamefont
  {Del~Duca}}, \bibinfo {author} {\bibfnamefont {C.}~\bibnamefont {Duhr}}, \
  and\ \bibinfo {author} {\bibfnamefont {V.~A.}\ \bibnamefont {Smirnov}},\
  }\href {\doibase 10.1007/JHEP07(2011)064} {\bibfield  {journal} {\bibinfo
  {journal} {JHEP}\ }\textbf {\bibinfo {volume} {07}},\ \bibinfo {pages} {064}
  (\bibinfo {year} {2011}{\natexlab{a}})},\ \Eprint
  {http://arxiv.org/abs/1105.1333} {arXiv:1105.1333 [hep-th]} \BibitemShut
  {NoStop}%
\bibitem [{\citenamefont {Del~Duca}\ \emph
  {et~al.}(2011{\natexlab{b}})\citenamefont {Del~Duca}, \citenamefont {Dixon},
  \citenamefont {Drummond}, \citenamefont {Duhr}, \citenamefont {Henn},\ and\
  \citenamefont {Smirnov}}]{DelDuca:2011wh}%
  \BibitemOpen
  \bibfield  {author} {\bibinfo {author} {\bibfnamefont {V.}~\bibnamefont
  {Del~Duca}}, \bibinfo {author} {\bibfnamefont {L.~J.}\ \bibnamefont {Dixon}},
  \bibinfo {author} {\bibfnamefont {J.~M.}\ \bibnamefont {Drummond}}, \bibinfo
  {author} {\bibfnamefont {C.}~\bibnamefont {Duhr}}, \bibinfo {author}
  {\bibfnamefont {J.~M.}\ \bibnamefont {Henn}}, \ and\ \bibinfo {author}
  {\bibfnamefont {V.~A.}\ \bibnamefont {Smirnov}},\ }\href {\doibase
  10.1103/PhysRevD.84.045017} {\bibfield  {journal} {\bibinfo  {journal} {Phys.
  Rev. D}\ }\textbf {\bibinfo {volume} {84}},\ \bibinfo {pages} {045017}
  (\bibinfo {year} {2011}{\natexlab{b}})},\ \Eprint
  {http://arxiv.org/abs/1105.2011} {arXiv:1105.2011 [hep-th]} \BibitemShut
  {NoStop}%
\bibitem [{\citenamefont {Arkani-Hamed}\ \emph {et~al.}(2019)\citenamefont
  {Arkani-Hamed}, \citenamefont {Lam},\ and\ \citenamefont
  {Spradlin}}]{Arkani-Hamed:2019rds}%
  \BibitemOpen
  \bibfield  {author} {\bibinfo {author} {\bibfnamefont {N.}~\bibnamefont
  {Arkani-Hamed}}, \bibinfo {author} {\bibfnamefont {T.}~\bibnamefont {Lam}}, \
  and\ \bibinfo {author} {\bibfnamefont {M.}~\bibnamefont {Spradlin}},\
  }\href@noop {} {\  (\bibinfo {year} {2019})},\ \Eprint
  {http://arxiv.org/abs/1912.08222} {arXiv:1912.08222 [hep-th]} \BibitemShut
  {NoStop}%
\bibitem [{\citenamefont {Henke}\ and\ \citenamefont
  {Papathanasiou}(2020)}]{Henke:2019hve}%
  \BibitemOpen
  \bibfield  {author} {\bibinfo {author} {\bibfnamefont {N.}~\bibnamefont
  {Henke}}\ and\ \bibinfo {author} {\bibfnamefont {G.}~\bibnamefont
  {Papathanasiou}},\ }\href {\doibase 10.1007/JHEP08(2020)005} {\bibfield
  {journal} {\bibinfo  {journal} {JHEP}\ }\textbf {\bibinfo {volume} {08}},\
  \bibinfo {pages} {005} (\bibinfo {year} {2020})},\ \Eprint
  {http://arxiv.org/abs/1912.08254} {arXiv:1912.08254 [hep-th]} \BibitemShut
  {NoStop}%
\bibitem [{\citenamefont {{J.M.~Drummond, J.~Foster, \"O.~G\"urdo\u gan,
  C.~Kalousios}}(2019)}]{Drummond:2019cxm}%
  \BibitemOpen
  \bibfield  {author} {\bibinfo {author} {\bibnamefont {{J.M.~Drummond,
  J.~Foster, \"O.~G\"urdo\u gan, C.~Kalousios}}},\ }\href@noop {} {\  (\bibinfo
  {year} {2019})},\ \Eprint {http://arxiv.org/abs/1912.08217} {arXiv:1912.08217
  [hep-th]} \BibitemShut {NoStop}%
\bibitem [{\citenamefont {Caron-Huot}(2011)}]{CaronHuot:2011ky}%
  \BibitemOpen
  \bibfield  {author} {\bibinfo {author} {\bibfnamefont {S.}~\bibnamefont
  {Caron-Huot}},\ }\href {\doibase 10.1007/JHEP12(2011)066} {\bibfield
  {journal} {\bibinfo  {journal} {JHEP}\ }\textbf {\bibinfo {volume} {12}},\
  \bibinfo {pages} {066} (\bibinfo {year} {2011})},\ \Eprint
  {http://arxiv.org/abs/1105.5606} {arXiv:1105.5606 [hep-th]} \BibitemShut
  {NoStop}%
\bibitem [{\citenamefont {Zhang}\ \emph {et~al.}(2019)\citenamefont {Zhang},
  \citenamefont {Li},\ and\ \citenamefont {He}}]{Zhang:2019vnm}%
  \BibitemOpen
  \bibfield  {author} {\bibinfo {author} {\bibfnamefont {C.}~\bibnamefont
  {Zhang}}, \bibinfo {author} {\bibfnamefont {Z.}~\bibnamefont {Li}}, \ and\
  \bibinfo {author} {\bibfnamefont {S.}~\bibnamefont {He}},\ }\href@noop {} {\
  (\bibinfo {year} {2019})},\ \Eprint {http://arxiv.org/abs/1911.01290}
  {arXiv:1911.01290 [hep-th]} \BibitemShut {NoStop}%
\bibitem [{\citenamefont {He}\ \emph {et~al.}(2020)\citenamefont {He},
  \citenamefont {Li},\ and\ \citenamefont {Zhang}}]{He:2020vob}%
  \BibitemOpen
  \bibfield  {author} {\bibinfo {author} {\bibfnamefont {S.}~\bibnamefont
  {He}}, \bibinfo {author} {\bibfnamefont {Z.}~\bibnamefont {Li}}, \ and\
  \bibinfo {author} {\bibfnamefont {C.}~\bibnamefont {Zhang}},\ }\href@noop {}
  {\  (\bibinfo {year} {2020})},\ \Eprint {http://arxiv.org/abs/2009.11471}
  {arXiv:2009.11471 [hep-th]} \BibitemShut {NoStop}%
\bibitem [{\citenamefont {Mago}\ \emph {et~al.}(2020)\citenamefont {Mago},
  \citenamefont {Schreiber}, \citenamefont {Spradlin},\ and\ \citenamefont
  {Volovich}}]{Mago:2020kmp}%
  \BibitemOpen
  \bibfield  {author} {\bibinfo {author} {\bibfnamefont {J.}~\bibnamefont
  {Mago}}, \bibinfo {author} {\bibfnamefont {A.}~\bibnamefont {Schreiber}},
  \bibinfo {author} {\bibfnamefont {M.}~\bibnamefont {Spradlin}}, \ and\
  \bibinfo {author} {\bibfnamefont {A.}~\bibnamefont {Volovich}},\ }\href
  {\doibase 10.1007/JHEP10(2020)128} {\bibfield  {journal} {\bibinfo  {journal}
  {JHEP}\ }\textbf {\bibinfo {volume} {10}},\ \bibinfo {pages} {128} (\bibinfo
  {year} {2020})},\ \Eprint {http://arxiv.org/abs/2007.00646} {arXiv:2007.00646
  [hep-th]} \BibitemShut {NoStop}%
\bibitem [{\citenamefont {He}\ and\ \citenamefont {Li}(2020)}]{He:2020uhb}%
  \BibitemOpen
  \bibfield  {author} {\bibinfo {author} {\bibfnamefont {S.}~\bibnamefont
  {He}}\ and\ \bibinfo {author} {\bibfnamefont {Z.}~\bibnamefont {Li}},\
  }\href@noop {} {\  (\bibinfo {year} {2020})},\ \Eprint
  {http://arxiv.org/abs/2007.01574} {arXiv:2007.01574 [hep-th]} \BibitemShut
  {NoStop}%
\bibitem [{\citenamefont {Hodges}(2013)}]{Hodges:2009hk}%
  \BibitemOpen
  \bibfield  {author} {\bibinfo {author} {\bibfnamefont {A.}~\bibnamefont
  {Hodges}},\ }\href {\doibase 10.1007/JHEP05(2013)135} {\bibfield  {journal}
  {\bibinfo  {journal} {Journal of High Energy Physics}\ }\textbf {\bibinfo
  {volume} {1305}},\ \bibinfo {pages} {135} (\bibinfo {year} {2013})},\ \Eprint
  {http://arxiv.org/abs/0905.1473} {arXiv:0905.1473 [hep-th]} \BibitemShut
  {NoStop}%
\bibitem [{\citenamefont {Spradlin}\ and\ \citenamefont
  {Volovich}(2011)}]{Spradlin:2011wp}%
  \BibitemOpen
  \bibfield  {author} {\bibinfo {author} {\bibfnamefont {M.}~\bibnamefont
  {Spradlin}}\ and\ \bibinfo {author} {\bibfnamefont {A.}~\bibnamefont
  {Volovich}},\ }\href {\doibase 10.1007/JHEP11(2011)084} {\bibfield  {journal}
  {\bibinfo  {journal} {JHEP}\ }\textbf {\bibinfo {volume} {11}},\ \bibinfo
  {pages} {084} (\bibinfo {year} {2011})},\ \Eprint
  {http://arxiv.org/abs/1105.2024} {arXiv:1105.2024 [hep-th]} \BibitemShut
  {NoStop}%
\bibitem [{\citenamefont {Abreu}\ \emph
  {et~al.}(2020{\natexlab{a}})\citenamefont {Abreu}, \citenamefont {Ita},
  \citenamefont {Moriello}, \citenamefont {Page}, \citenamefont {Tschernow},\
  and\ \citenamefont {Zeng}}]{Abreu:2020jxa}%
  \BibitemOpen
  \bibfield  {author} {\bibinfo {author} {\bibfnamefont {S.}~\bibnamefont
  {Abreu}}, \bibinfo {author} {\bibfnamefont {H.}~\bibnamefont {Ita}}, \bibinfo
  {author} {\bibfnamefont {F.}~\bibnamefont {Moriello}}, \bibinfo {author}
  {\bibfnamefont {B.}~\bibnamefont {Page}}, \bibinfo {author} {\bibfnamefont
  {W.}~\bibnamefont {Tschernow}}, \ and\ \bibinfo {author} {\bibfnamefont
  {M.}~\bibnamefont {Zeng}},\ }\href@noop {} {\  (\bibinfo {year}
  {2020}{\natexlab{a}})},\ \Eprint {http://arxiv.org/abs/2005.04195}
  {arXiv:2005.04195 [hep-ph]} \BibitemShut {NoStop}%
\bibitem [{\citenamefont {Gehrmann}\ \emph {et~al.}(2016)\citenamefont
  {Gehrmann}, \citenamefont {Henn},\ and\ \citenamefont
  {Lo~Presti}}]{Gehrmann:2015bfy}%
  \BibitemOpen
  \bibfield  {author} {\bibinfo {author} {\bibfnamefont {T.}~\bibnamefont
  {Gehrmann}}, \bibinfo {author} {\bibfnamefont {J.}~\bibnamefont {Henn}}, \
  and\ \bibinfo {author} {\bibfnamefont {N.}~\bibnamefont {Lo~Presti}},\ }\href
  {\doibase 10.1103/PhysRevLett.116.062001} {\bibfield  {journal} {\bibinfo
  {journal} {Phys. Rev. Lett.}\ }\textbf {\bibinfo {volume} {116}},\ \bibinfo
  {pages} {062001} (\bibinfo {year} {2016})},\ \bibinfo {note} {[Erratum:
  Phys.Rev.Lett. 116, 189903 (2016)]},\ \Eprint
  {http://arxiv.org/abs/1511.05409} {arXiv:1511.05409 [hep-ph]} \BibitemShut
  {NoStop}%
\bibitem [{\citenamefont {Chicherin}\ \emph {et~al.}(2018)\citenamefont
  {Chicherin}, \citenamefont {Henn},\ and\ \citenamefont
  {Mitev}}]{Chicherin:2017dob}%
  \BibitemOpen
  \bibfield  {author} {\bibinfo {author} {\bibfnamefont {D.}~\bibnamefont
  {Chicherin}}, \bibinfo {author} {\bibfnamefont {J.}~\bibnamefont {Henn}}, \
  and\ \bibinfo {author} {\bibfnamefont {V.}~\bibnamefont {Mitev}},\ }\href
  {\doibase 10.1007/JHEP05(2018)164} {\bibfield  {journal} {\bibinfo  {journal}
  {JHEP}\ }\textbf {\bibinfo {volume} {05}},\ \bibinfo {pages} {164} (\bibinfo
  {year} {2018})},\ \Eprint {http://arxiv.org/abs/1712.09610} {arXiv:1712.09610
  [hep-th]} \BibitemShut {NoStop}%
\bibitem [{\citenamefont {Abreu}\ \emph
  {et~al.}(2019{\natexlab{a}})\citenamefont {Abreu}, \citenamefont {Dixon},
  \citenamefont {Herrmann}, \citenamefont {Page},\ and\ \citenamefont
  {Zeng}}]{Abreu:2018aqd}%
  \BibitemOpen
  \bibfield  {author} {\bibinfo {author} {\bibfnamefont {S.}~\bibnamefont
  {Abreu}}, \bibinfo {author} {\bibfnamefont {L.~J.}\ \bibnamefont {Dixon}},
  \bibinfo {author} {\bibfnamefont {E.}~\bibnamefont {Herrmann}}, \bibinfo
  {author} {\bibfnamefont {B.}~\bibnamefont {Page}}, \ and\ \bibinfo {author}
  {\bibfnamefont {M.}~\bibnamefont {Zeng}},\ }\href {\doibase
  10.1103/PhysRevLett.122.121603} {\bibfield  {journal} {\bibinfo  {journal}
  {Phys. Rev. Lett.}\ }\textbf {\bibinfo {volume} {122}},\ \bibinfo {pages}
  {121603} (\bibinfo {year} {2019}{\natexlab{a}})},\ \Eprint
  {http://arxiv.org/abs/1812.08941} {arXiv:1812.08941 [hep-th]} \BibitemShut
  {NoStop}%
\bibitem [{\citenamefont {Chicherin}\ \emph
  {et~al.}(2019{\natexlab{a}})\citenamefont {Chicherin}, \citenamefont
  {Gehrmann}, \citenamefont {Henn}, \citenamefont {Wasser}, \citenamefont
  {Zhang},\ and\ \citenamefont {Zoia}}]{Chicherin:2018yne}%
  \BibitemOpen
  \bibfield  {author} {\bibinfo {author} {\bibfnamefont {D.}~\bibnamefont
  {Chicherin}}, \bibinfo {author} {\bibfnamefont {T.}~\bibnamefont {Gehrmann}},
  \bibinfo {author} {\bibfnamefont {J.~M.}\ \bibnamefont {Henn}}, \bibinfo
  {author} {\bibfnamefont {P.}~\bibnamefont {Wasser}}, \bibinfo {author}
  {\bibfnamefont {Y.}~\bibnamefont {Zhang}}, \ and\ \bibinfo {author}
  {\bibfnamefont {S.}~\bibnamefont {Zoia}},\ }\href {\doibase
  10.1103/PhysRevLett.122.121602} {\bibfield  {journal} {\bibinfo  {journal}
  {Phys. Rev. Lett.}\ }\textbf {\bibinfo {volume} {122}},\ \bibinfo {pages}
  {121602} (\bibinfo {year} {2019}{\natexlab{a}})},\ \Eprint
  {http://arxiv.org/abs/1812.11057} {arXiv:1812.11057 [hep-th]} \BibitemShut
  {NoStop}%
\bibitem [{\citenamefont {Chicherin}\ \emph
  {et~al.}(2019{\natexlab{b}})\citenamefont {Chicherin}, \citenamefont
  {Gehrmann}, \citenamefont {Henn}, \citenamefont {Wasser}, \citenamefont
  {Zhang},\ and\ \citenamefont {Zoia}}]{Chicherin:2019xeg}%
  \BibitemOpen
  \bibfield  {author} {\bibinfo {author} {\bibfnamefont {D.}~\bibnamefont
  {Chicherin}}, \bibinfo {author} {\bibfnamefont {T.}~\bibnamefont {Gehrmann}},
  \bibinfo {author} {\bibfnamefont {J.~M.}\ \bibnamefont {Henn}}, \bibinfo
  {author} {\bibfnamefont {P.}~\bibnamefont {Wasser}}, \bibinfo {author}
  {\bibfnamefont {Y.}~\bibnamefont {Zhang}}, \ and\ \bibinfo {author}
  {\bibfnamefont {S.}~\bibnamefont {Zoia}},\ }\href {\doibase
  10.1007/JHEP03(2019)115} {\bibfield  {journal} {\bibinfo  {journal} {JHEP}\
  }\textbf {\bibinfo {volume} {03}},\ \bibinfo {pages} {115} (\bibinfo {year}
  {2019}{\natexlab{b}})},\ \Eprint {http://arxiv.org/abs/1901.05932}
  {arXiv:1901.05932 [hep-th]} \BibitemShut {NoStop}%
\bibitem [{\citenamefont {Abreu}\ \emph
  {et~al.}(2019{\natexlab{b}})\citenamefont {Abreu}, \citenamefont {Dixon},
  \citenamefont {Herrmann}, \citenamefont {Page},\ and\ \citenamefont
  {Zeng}}]{Abreu:2019rpt}%
  \BibitemOpen
  \bibfield  {author} {\bibinfo {author} {\bibfnamefont {S.}~\bibnamefont
  {Abreu}}, \bibinfo {author} {\bibfnamefont {L.~J.}\ \bibnamefont {Dixon}},
  \bibinfo {author} {\bibfnamefont {E.}~\bibnamefont {Herrmann}}, \bibinfo
  {author} {\bibfnamefont {B.}~\bibnamefont {Page}}, \ and\ \bibinfo {author}
  {\bibfnamefont {M.}~\bibnamefont {Zeng}},\ }\href {\doibase
  10.1007/JHEP03(2019)123} {\bibfield  {journal} {\bibinfo  {journal} {JHEP}\
  }\textbf {\bibinfo {volume} {03}},\ \bibinfo {pages} {123} (\bibinfo {year}
  {2019}{\natexlab{b}})},\ \Eprint {http://arxiv.org/abs/1901.08563}
  {arXiv:1901.08563 [hep-th]} \BibitemShut {NoStop}%
\bibitem [{\citenamefont {Abreu}\ \emph
  {et~al.}(2020{\natexlab{b}})\citenamefont {Abreu}, \citenamefont {Page},
  \citenamefont {Pascual},\ and\ \citenamefont {Sotnikov}}]{Abreu:2020cwb}%
  \BibitemOpen
  \bibfield  {author} {\bibinfo {author} {\bibfnamefont {S.}~\bibnamefont
  {Abreu}}, \bibinfo {author} {\bibfnamefont {B.}~\bibnamefont {Page}},
  \bibinfo {author} {\bibfnamefont {E.}~\bibnamefont {Pascual}}, \ and\
  \bibinfo {author} {\bibfnamefont {V.}~\bibnamefont {Sotnikov}},\ }\href@noop
  {} {\  (\bibinfo {year} {2020}{\natexlab{b}})},\ \Eprint
  {http://arxiv.org/abs/2010.15834} {arXiv:2010.15834 [hep-ph]} \BibitemShut
  {NoStop}%
\bibitem [{\citenamefont {Sotnikov}(2020)}]{Vasilycomm}%
  \BibitemOpen
  \bibfield  {author} {\bibinfo {author} {\bibfnamefont {V.}~\bibnamefont
  {Sotnikov}},\ }\href@noop {} {}\bibinfo {howpublished} {private
  communication} (\bibinfo {year} {2020})\BibitemShut {NoStop}%
\bibitem [{\citenamefont {Badger}\ \emph {et~al.}(2019)\citenamefont {Badger},
  \citenamefont {Br\o{}nnum-Hansen}, \citenamefont {Hartanto},\ and\
  \citenamefont {Peraro}}]{Badger:2018enw}%
  \BibitemOpen
  \bibfield  {author} {\bibinfo {author} {\bibfnamefont {S.}~\bibnamefont
  {Badger}}, \bibinfo {author} {\bibfnamefont {C.}~\bibnamefont
  {Br\o{}nnum-Hansen}}, \bibinfo {author} {\bibfnamefont {H.~B.}\ \bibnamefont
  {Hartanto}}, \ and\ \bibinfo {author} {\bibfnamefont {T.}~\bibnamefont
  {Peraro}},\ }\href {\doibase 10.1007/JHEP01(2019)186} {\bibfield  {journal}
  {\bibinfo  {journal} {JHEP}\ }\textbf {\bibinfo {volume} {01}},\ \bibinfo
  {pages} {186} (\bibinfo {year} {2019})},\ \Eprint
  {http://arxiv.org/abs/1811.11699} {arXiv:1811.11699 [hep-ph]} \BibitemShut
  {NoStop}%
\bibitem [{\citenamefont {Abreu}\ \emph
  {et~al.}(2019{\natexlab{c}})\citenamefont {Abreu}, \citenamefont {Dormans},
  \citenamefont {Febres~Cordero}, \citenamefont {Ita}, \citenamefont {Page},\
  and\ \citenamefont {Sotnikov}}]{Abreu:2019odu}%
  \BibitemOpen
  \bibfield  {author} {\bibinfo {author} {\bibfnamefont {S.}~\bibnamefont
  {Abreu}}, \bibinfo {author} {\bibfnamefont {J.}~\bibnamefont {Dormans}},
  \bibinfo {author} {\bibfnamefont {F.}~\bibnamefont {Febres~Cordero}},
  \bibinfo {author} {\bibfnamefont {H.}~\bibnamefont {Ita}}, \bibinfo {author}
  {\bibfnamefont {B.}~\bibnamefont {Page}}, \ and\ \bibinfo {author}
  {\bibfnamefont {V.}~\bibnamefont {Sotnikov}},\ }\href {\doibase
  10.1007/JHEP05(2019)084} {\bibfield  {journal} {\bibinfo  {journal} {JHEP}\
  }\textbf {\bibinfo {volume} {05}},\ \bibinfo {pages} {084} (\bibinfo {year}
  {2019}{\natexlab{c}})},\ \Eprint {http://arxiv.org/abs/1904.00945}
  {arXiv:1904.00945 [hep-ph]} \BibitemShut {NoStop}%
\bibitem [{\citenamefont {Badger}\ and\ \citenamefont
  {Zoia}(2020)}]{Simonecomm}%
  \BibitemOpen
  \bibfield  {author} {\bibinfo {author} {\bibfnamefont {S.}~\bibnamefont
  {Badger}}\ and\ \bibinfo {author} {\bibfnamefont {S.}~\bibnamefont {Zoia}},\
  }\href@noop {} {}\bibinfo {howpublished} {private communication} (\bibinfo
  {year} {2020})\BibitemShut {NoStop}%
\bibitem [{\citenamefont {Brandhuber}\ \emph {et~al.}(2012)\citenamefont
  {Brandhuber}, \citenamefont {Travaglini},\ and\ \citenamefont
  {Yang}}]{Brandhuber:2012vm}%
  \BibitemOpen
  \bibfield  {author} {\bibinfo {author} {\bibfnamefont {A.}~\bibnamefont
  {Brandhuber}}, \bibinfo {author} {\bibfnamefont {G.}~\bibnamefont
  {Travaglini}}, \ and\ \bibinfo {author} {\bibfnamefont {G.}~\bibnamefont
  {Yang}},\ }\href {\doibase 10.1007/JHEP05(2012)082} {\bibfield  {journal}
  {\bibinfo  {journal} {JHEP}\ }\textbf {\bibinfo {volume} {05}},\ \bibinfo
  {pages} {082} (\bibinfo {year} {2012})},\ \Eprint
  {http://arxiv.org/abs/1201.4170} {arXiv:1201.4170 [hep-th]} \BibitemShut
  {NoStop}%
\bibitem [{\citenamefont {Arkani-Hamed}\ \emph {et~al.}(2020)\citenamefont
  {Arkani-Hamed}, \citenamefont {He},\ and\ \citenamefont
  {Lam}}]{Arkani-Hamed:2020tuz}%
  \BibitemOpen
  \bibfield  {author} {\bibinfo {author} {\bibfnamefont {N.}~\bibnamefont
  {Arkani-Hamed}}, \bibinfo {author} {\bibfnamefont {S.}~\bibnamefont {He}}, \
  and\ \bibinfo {author} {\bibfnamefont {T.}~\bibnamefont {Lam}},\ }\href@noop
  {} {\  (\bibinfo {year} {2020})},\ \Eprint {http://arxiv.org/abs/2005.11419}
  {arXiv:2005.11419 [math.AG]} \BibitemShut {NoStop}%
\end{thebibliography}%

\end{document}